%% file: main.tex
\documentclass[]{resources/aa}
\usepackage{graphicx}
\usepackage[varg]{txfonts}
\usepackage{physics}
\usepackage{amsmath, amssymb}
\usepackage{dsfont}
\usepackage{bbm}
\usepackage{xcolor}
\usepackage{ulem}
\usepackage{cancel}   % MG for \cancel{}  
\usepackage{soul}     % MG  for \st{}
\usepackage{xpatch}
\usepackage{hyperref}
\usepackage{import}
\usepackage{makeidx}
\usepackage{changes}
\usepackage{svg}
\usepackage{adjustbox}
\usepackage[flushleft]{threeparttable}
\usepackage{placeins}

\newcommand{\sref}[1]{Sec.~\ref{#1}}
\newcommand{\tab}[1]{Table~\ref{#1}}
\newcommand{\fig}[1]{Fig.~\ref{#1}}
\newcommand{\equ}[1]{Eq.~(\ref{#1})}
\newcommand{\equo}[1]{Eq.~\ref{#1}}
\newcommand{\equs}[2]{Eqs.~(\ref{#1})~-~(\ref{#2})}
\newcommand{\Msolpyr}{\mathrm{M_\odot~yr}^{-1}}

\newcommand{\colout}[1]{\bgroup\markoverwith{\textcolor{#1}{\rule[.5ex]{2pt}{0.4pt}}}\ULon}

\makeatletter
\renewcommand*\aa@pageof{, page \thepage{} of \pageref*{LastPage}}

\makeatother

\makeindex

\begin{document}

\import{sections/}{header}

\import{sections/}{introduction}

\import{sections/}{model_description}

% \import{sections/}{spin_model}

\import{sections/}{results}

% \import{sections/}{discussion}

\import{sections/}{conclusion}

\import{sections/}{acknowledgments}

% ------------------------------------------------------------------
% FOOTER
% ------------------------------------------------------------------
\import{}{footer}

\appendix
\import{sections/}{appendix}

\end{document}

%% file: sections/header.tex
\titlerunning{The primordial (post-disk) spin distribution of M dwarf stars}
\authorrunning{L. Gehrig et al.}
\title{The post-disk (or primordial) spin distribution of M dwarf stars} 
% \title{On the angular momentum distribution of young M dwarfs after the disk phase} 

% alternative title: The post-disk (or primordial) spin distribution of M dwarf stars

% \subtitle{....}

\author{
L.~Gehrig\inst{1},
E.~Gaidos\inst{1,2},
M.~G\"udel\inst{1,3}
}

\institute{
 Department of Astrophysics, University of Vienna,
 Türkenschanzstrasse 17, A-1180 Vienna, Austria
 \label{1}
\and
 Department of Earth Sciences, University of Hawai'i at M\={a}noa,
 Honolulu, HI 96822 USA
 \label{2}
\and
  Max-Planck-Institut f\"ur Astronomie, K\"onigstuhl 17, 69117 Heidelberg, Germany
 \label{3}}
%\today
\date{Received ....; accepted ....}

\abstract
{The rotation periods of young low-mass stars after disks have dissipated ($\lesssim$10 Myr) but before magnetized winds have removed significant angular momentum is an important launch point for gyrochronology and models of stellar rotational evolution; the rotation of these stars also regulates the magnetic activity and the intensity of high-energy emission that affects any close-in planets. A recent analysis of young M dwarf stars suggests a distribution of specific angular momentum (SAM) that is mass-independent, but the physical basis of this observation is unclear.}
{We investigate the influence of an accretion disk on the angular momentum (AM) evolution of young M dwarfs, which parameters govern the AM distribution after the disk phase, and whether this leads to a mass-independent distribution of SAM.}
{We use a combination of protostellar spin and implicit hydrodynamic disk evolution models to model the innermost disk ($\sim 0.01$~AU), including a self-consistent calculation of the accretion rate onto the star, non-Keplerian disk rotation, and the influence of stellar magnetic torques over the entire disk lifetime.  We execute and analyze over 500 long-term simulations of the combined stellar and disk evolution.}
{We find that above an initial rate $\Dot{M}_\mathrm{crit} \sim 10^{-8}~\Msolpyr$ accretion "erases" the initial SAM of M dwarfs during the disk lifetime, and stellar rotation converges to values of SAM that are largely independent of initial conditions.  For stellar masses $> 0.3~\mathrm{M_\odot}$, we find that observed initial accretion rates $\Dot{M}_\mathrm{init}$ are comparable to or exceed $\Dot{M}_\mathrm{crit}$. Furthermore, stellar SAM after the disk phase scales with the stellar magnetic field strength as a power-law with an exponent of $-1.1$.
For lower stellar masses, $\Dot{M}_\mathrm{init}$ is predicted to be smaller than $\Dot{M}_\mathrm{crit}$ and the initial conditions are  imprinted in the stellar SAM after the disk phase.}
{To explain the observed mass-independent distribution of SAM, the stellar magnetic field strength has to range between 20~G and 500~G (700~G and 1500~G) for a 0.1~$\mathrm{M_\odot}$ (0.6~$\mathrm{M_\odot}$) star. 
These values match observed large-scale magnetic field measurements of young M~dwarfs and the positive relation between stellar mass and magnetic field strength agrees with a  theoretically-motivated scaling relation.
The scaling law between stellar SAM, mass, and the magnetic field strength is consistent for young stars, where these parameters are constrained by observations.
Due to the very limited number of available data, we advocate for efforts to obtain more such measurements.
Our results provide new constraints on the relation between stellar mass and magnetic field strength and can be used as initial conditions for future stellar spin models, starting after the disk phase.
}


\maketitle

%% file: sections/introduction.tex
\section{Introduction}
\label{sec:intro}

One of the major challenges in stellar astrophysics is the origin and evolution of stellar angular momentum (AM) and spin. Over the last decade, observations of young clusters have yielded new insight into the evolution of pre-main sequence (PMS) and young ($\lesssim1$Gyr) main sequence dwarf stars \citep[e.g.,][]{herbst02,Irwin08,Irwin09, Hartmann09,Meibom11,Affer13,Gallet13}.  Such co-eval groups of stars with established ages and a range of masses provide benchmarks for understanding how that rotational evolution varies with stellar mass and age.  The spin evolution of low-mass stars (starting from the Class~II phase) can be divided into three phases \citep{Gallet13}: (1) exchange of angular momentum between a young stellar object (YSO) and its primordial disk; (2) simple contraction and spin-up of the PMS star after the disk has dissipated; (3) gradual spin-down on the main sequence by loss of AM through a magnetized wind.  

Understanding the initial rotation distribution and subsequent evolution of young stars is important to relating stellar rotation to ages \citep["gyrochronology";][]{Barnes07}.  Furthermore, rotation drives magneto dynamo operation in stars with convective envelopes, and magnetic activity is responsible for heating the upper stellar atmosphere, which in terms is the primary source of X-rays and far- and extreme-ultraviolet (XUV) emission from low-mass stars \citep[e.g.,][]{Pallavicini81, Micela85, Pizzolato03,Wright11, France18}.  XUV irradiation by the host star will heat and inflate the upper atmospheres of close-in planets and can lead to evaporation and even complete loss of the atmosphere \citep[e.g.,][]{Guedel97, Lammer03, Cecchi09, Lecavelier10, Sanz11, France13, Nortmann18}. XUV-driven loss of water as atomic H can affect the habitability of rocky planets \citep[e.g.,][]{Buccino06}.

These issues come to the fore for studies of M~dwarf stars ($M_\star \approx 0.1 - 0.6~\mathrm{M_\odot}$) as hosts of planets, particularly rocky and potentially habitable planets.
M dwarfs are the most common type of star \citep[e.g.,][]{Chabrier03} and the small size and low mass of M~dwarfs favor the detection of planets by the transit and radial velocity methods on close-in orbits. In addition, the lower luminosities of M~dwarfs mean that their ``habitable zones", where the equilibrium temperatures of planets would permit surface liquid water, are much closer to the stars \citep[e.g.,][]{Zechmeister09}.
However, these stars are intrinsically faint and more difficult to study at the distances of benchmark clusters and as a result, studies and modeling of their rotational evolution and establishment of gyrochronological benchmarks are at an early stage.

Robust measurements of rotation periods for M dwarfs in a few star-forming regions and young clusters have been provided by the precise photometry of the \textit{Kepler} space telescope, re-purposed to observe multiple ecliptic fields in the two-wheel K2 mission \citep{Howell2014}.  These groups include the Taurus \citep[1-5 Myr;][]{Rebull2020}, $\rho$ Ophiuchus (3-5 Myr), and Upper Scorpius \citep[$\sim$10 Myr,][]{Rebull2018} star-forming regions, and the Pleiades \citep[$\approx$125 Myr][]{Rebull2016}, Hyades \citep[$\approx$650 Myr][]{Douglas2019}, and Praesepe \citep[$\approx$800][]{Rebull2017,Douglas2017} clusters.  \citet{Somers17} showed that the distributions of \textit{specific} (per unit mass) angular momentum (SAM) of M~dwarfs in Upper Scorpius and the Pleiades were similar, approximately independent of stellar mass, but with a significant scatter.  Since by the age of Upper Scorpius circumstellar disks have largely (but not completely) dissipated \citep{Mamajek09}, the similarity of the two distributions means that angular momentum  loss through winds during the first $\le$125 Myr is $\lesssim$40\%.\footnote{Note that contracting PMS stars spin up so the distributions of \emph{rotation periods} differ.}   It also follows that the origin of the mass-independent SAM distribution lies in the earlier YSO phase and could be a product of the interaction between stars and their disks.

Our study models the YSO phase \citep[$\lesssim 10$~Myr, e.g.,][]{Mamajek09} of single M dwarfs to determine if star-disk interactions can explain the observed distribution of SAM and rotation at post-disk times.  
While the rotational evolution of young, disk-free M~dwarfs on the PMS has been previously studied \citep[e.g.,][]{Bouvier14, Matt15}, the influence of disks on their rotation is less well explored and most of the work has consisted of long-term spin evolution calculations \citep[e.g.,][]{Armitage96, Matt10, Matt12, Gallet19}.
During this phase, mass is accreted from the disk onto the star, altering the star's angular momentum and subsequent rotational evolution.  It has been proposed that this exchange leads to the star rotating at the orbital period of the disk inner edge \citep[``disk-locking", e.g.,][]{Ghosh79, Koenigl91}. 
Calculations by \cite{Armitage96} suggest that massive disks coupled to the stellar magnetic field can regulate a star's spin evolution while low mass disks fail to slow down the star as it contracts and spins up. While \citet{Armitage96} include simplified time-dependent disk evolution in their model, they assume the entire disk couples to the stellar magnetic field lines.
However, due to the limited strength of the coupling between the disk and the stellar magnetic field lines the disk may not be able to remove the AM required to balance spin-up due to stellar accretion and stellar contraction and keep the star's rotation "locked" to the disk \citep[e.g.,][]{Matt05b, Zanni13}.  Moreover, disk-locking does not explain the existence of very slowly rotating stars \citep[e.g.,][]{Matt05}.

Other phenomena may be in play during the disk phase:  Accretion-powered outflows or winds could also remove angular momentum from the star \citep[APSW, e.g.,][]{Matt05, Matt08, Matt08b,Matt12,Finnley18, Gallet19}. Magnetized disk winds  can remove mass and angular momentum from the disk before these are accreted to the star  \citep[e.g.,][]{Shu94, Ferreira00, Romanova09, Zanni13}.  Magnetospheric ejections, in which the stellar and/or disk magnetic field is loaded with mass from the disk \citep[]{Zanni13}, can also transport angular momentum between the star and disk.  

Self-consistent modeling of the accretion rate in the disk $\Dot{M}_\mathrm{disk}$ and onto the star is important to robustly describe mass and angular momentum exchange. Previous studies fixed the accretion rate to a power-law or an exponential decay \citep[e.g.,][]{Matt10,Gallet19} and did not account for feedback between the star and disk.  Modeling the disk's inner edge ($\lesssim$0.1~au) is numerically challenging because of the short time steps that are required, but this can be achieved with implicit hydrodynamic codes which circumvent the time-step limitations of explicit codes, making long-term simulations feasible \citep[][]{Courant28, ragossnig20,Steiner21}.         

Here we combine a model of pre-main sequence stellar evolution \citep{Matt10} with an implicit hydrodynamic model of disk evolution \citep[TAPIR;][]{ragossnig20, Steiner21} that can more realistically and self-consistently calculate disk accretion, the effects of stellar magnetic torques and local pressure gradients, the position of the inner disk boundary, and the effect of the star on the disk.  We seek to understand (1) how the parameters describing the star-disk interaction govern the final distribution of SAM when disks have dissipated; and (2) whether the observed, mass-independent distribution of SAM is a natural outcome of our model.

%% file: sections/model_description.tex
\section{Model description}
\label{sec:model_description}

\begin{figure}
    \centering
    \resizebox{\hsize}{!}{\includegraphics[trim={3cm 1cm 10cm 0.3cm},clip]{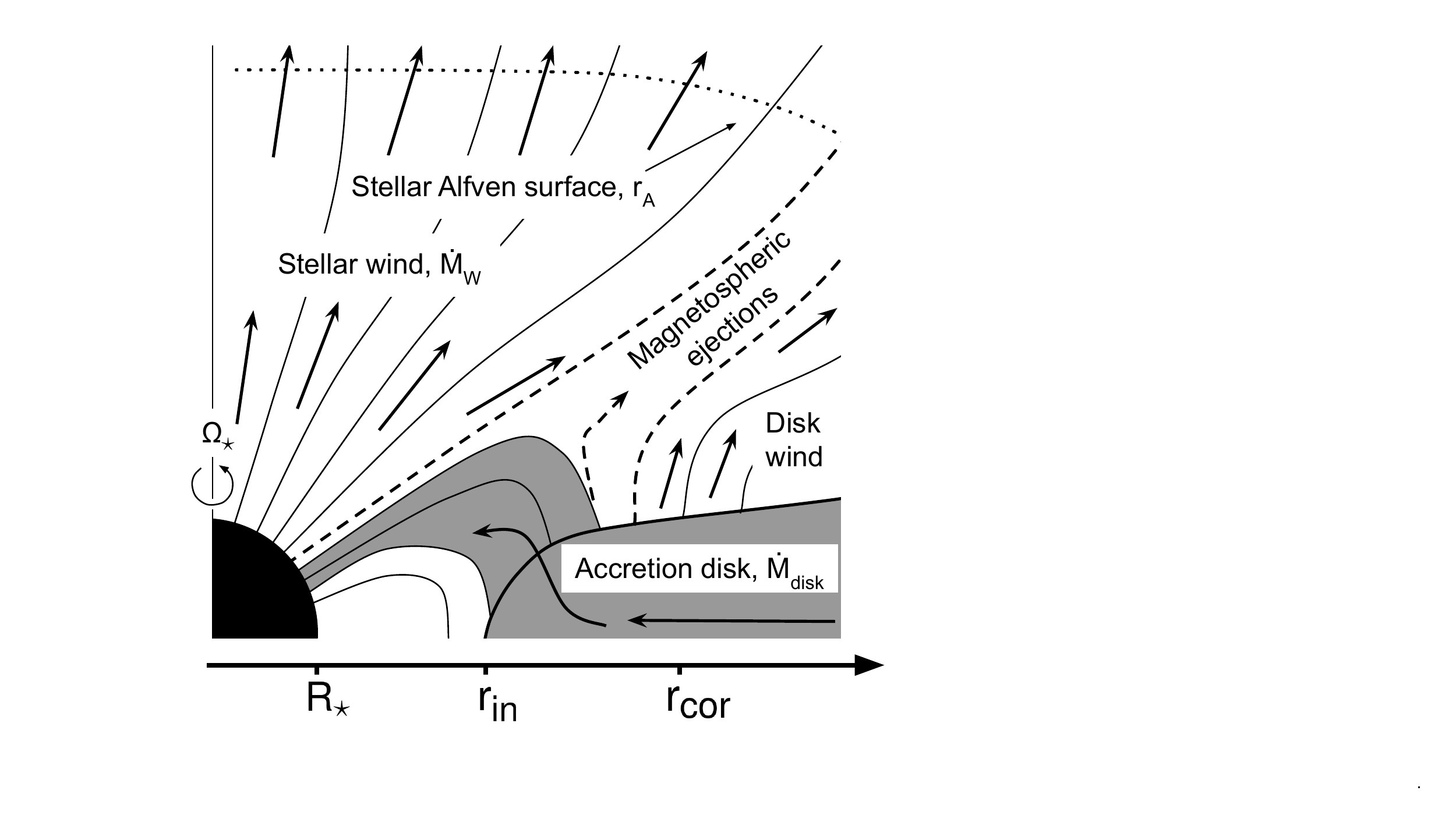}}

    \caption{Schematic cartoon of inner disk-star connection. The star (black quarter circle on the left) has a radius of $R_\star$ and rotates with $\Omega_\star$. Disk material (grey area) flows from the outer disk towards the inner disk radius $r_\mathrm{in}$, where it is forced into funnel flows and is accreted onto the star \citep[e.g.,][]{Romanova04, bessolaz08, Steiner21} with an accretion rate of $\dot{M}_\mathrm{disk}$. The accreted disk material increases the stellar angular momentum by $\Gamma_\mathrm{acc}$. Based on the model, a specific fraction of the accreted mass is driven into a stellar wind, which removes mass and angular momentum from the star ($\dot{M}_W$ and $\Gamma_\mathrm{W}$, respectively). The Alfven radius $r_\mathrm{A}$ acts as a lever arm for this stellar outflow. We note that this cartoon is not to scale and we can usually assume $r_\mathrm{A}\gg R_\star$. Angular momentum can be transferred towards or away from the star ($\Gamma_\mathrm{ME}$), depending on the position of $r_\mathrm{in}$ with respect to the co-rotation radius $r_\mathrm{cor}$ \citep[e.g.,][]{Zanni13}. Finally, magneto-centrifugally driven disk winds remove angular momentum and mass from the disk before it can be accreted on the star \citep[e.g.,][]{Konigl11, Bai2013}. We note that disk winds are not included in our model.}
    \label{fig:cartoon}
\end{figure}

We combine a model of stellar spin evolution  \citep[e.g.,][]{Matt10, Matt12, Gallet19} with a hydrodynamic description of a protostellar accretion disk. The model used in this study is already described in full detail in \cite{ragossnig20} and \cite{Steiner21} and we refer the reader to these papers for further information. The most important aspects, as well as differences from the version used in \cite{Steiner21}, are highlighted below.  Figure \ref{fig:cartoon} schematically illustrates the major features and important parameters of the model.

\subsection{Hydrodynamic disk model}\label{sec:hydrodynamic_disk_evolution}

Our simulations model the protoplanetary disk as a time-dependent, viscous accretion disk \citep[e.g.,][]{Shakura1973, armitage01}. Accretion rates are computed self-consistently and are not fixed. 
We use the one-dimensional implicit TAPIR code \citep[][]{ragossnig20, Steiner21}, which solves the equations for the surface density $\Sigma$, internal energy $e$ and the radial velocity $u_\mathrm{r}$ as well as the azimuthal velocity $u_\mathrm{\varphi}$ \citep[][]{ragossnig20, Steiner21}. Additionally, the adaptive grid equation introduced by \cite{Dorfi1987} is solved to ensure sufficient grid resolution and enable moving inner and outer disk boundaries. The accretion disk is assumed to be axisymmetric and in hydrostatic equilibrium in the vertical direction, and thus quasi 2-d. The model allows the velocity to deviate from the Keplerian azimuthal velocity $u_\mathrm{\varphi} \sim r^{-1/2}$, and effects such as the influence of a stellar magnetic field can be included, being of particular importance in the inner part of the disk \citep[e.g.,][]{Romanova02, bessolaz08}. Models based on the diffusion approximation \citep[e.g.,][]{pringle81, armitage01, zhu07, zhu10a} are not able to include these effects over the whole disk lifetime of $\lesssim 10$~Myr. The viscosity model used in this study is based on \cite{Shakura1973} with the kinematic viscosity $\nu = \alpha c_\mathrm{S} H_\mathrm{P}$. Here, $\alpha$, $c_\mathrm{S}$ and $H_\mathrm{P}$ denote the dimensionless viscosity parameter \citep[assumed to range between 0.005 and 0.01, e.g.,][]{zhu07,Vorobyov09,Yang18}, the speed of sound, and the vertical pressure scale height, respectively. 

In the inner disk, the stellar magnetic field acts on the partially ionized disk, and the motion of the disk alters the ambient magnetic field.  The magnetic field of the star is assumed to be co-rotating with the star and described as a dipole aligned with the rotation axis and perpendicular to the disk plane. However, the Keplerian motion of the disk tends to wind up locally vertical field lines and generates an azimuthal component of the magnetic field $B_\mathrm{\varphi}$ \citep[e.g.,][]{rappaport04, kluzniak07, Steiner21}.  The motion of ionized disk gas with respect to the local magnetic field induces a torque in the azimuthal direction and thus leads to an acceleration in the radial direction \citep[e.g.,][]{Romanova02}. The resulting drop in disk surface density towards the star \citep[e.g.,][]{bessolaz08, Steiner21} creates a pressure gradient.  The torque, induced by the stellar magnetic field, changes sign at the co-rotation radius
\begin{alignat}{2}
    &r_\mathrm{cor} &&= \left( \frac{G \, M_\star \, P_\star^2}{4 \, \pi^2} \right)^{1/3} \label{eq:corotation_radius} \;,
\end{alignat}
with $G$ being the gravitational constant, $M_\star$ the stellar mass and $P_\star$ the stellar rotational period.  Interior to $r_\mathrm{cor}$, the disk gas is decelerated in the azimuthal and accelerated in the radial direction towards the star. Exterior to $r_\mathrm{cor}$, the disk material is accelerated in the azimuthal direction, pushing it away from the star.  Towards the innermost part of the disk, the magnetic pressure ($\propto B_\star^2 (R_\star/r)^6$) overwhelms the gas pressure as well as the ram pressure of the disk material moving towards the star. At a so-called truncation radius $r_\mathrm{trunc}$, the disk material is forced along magnetic field lines and is ``funneled" towards the star \citep[e.g.,][]{Romanova02, Romanova14}.  
We choose $r_\mathrm{trunc}$ to be the inner radius of the accretion disk $r_\mathrm{in}$; this radius is calculated by equating the stellar magnetic pressure to the disk's ram pressure or gas pressure at the inner boundary, whichever is higher \citep[see][]{Steiner21}.
Similar to \cite{Romanova02} and \cite{Steiner21}, the inner boundary is treated as a ``free''/~Neumann boundary  (i.e., $\partial_\mathrm{r} \Sigma = 0$). At the outer boundary, the azimuthal velocity $u_\mathrm{\varphi}$ equals the Keplerian velocity, the radial velocity $u_\mathrm{r}$ is set to zero and the surface density $\Sigma$ as well as the internal energy $e$ are again treated with Neumann boundary conditions.
The outer disk radius is set to the position where the disk surface density has declined to $\Sigma_\mathrm{out} = 1~\mathrm{g/cm^2}$ \citep[similar to ][]{Vorobyov20}.

\subsection{Stellar spin model}\label{sec:stellar_spin_model}

The stellar spin evolution model used in this study is based on the work of \cite{Matt10} and \cite{Gallet19}. The simulations are started from a stellar age of $t_\mathrm{0}=1$~Myr \citep[similar to][]{Gallet19}, an age corresponding approximately to the beginning of the Class II YSO phase when the only significant accretion is through a disk. The star is considered to be fully convective and to rotate as a solid body. The stellar angular momentum can be written as $J_\star = I_\star \Omega_\star$, with a moment of inertia $I_\star = k^2 M_\star R_\star^2$ including the stellar radius $R_\star$ as well as the dimensionless radius of gyration $k$ and the stellar angular velocity $\Omega_\star$. We assume $k^2 = 0.2$ to be constant in time, which is a reasonable assumption over the first 10~Myr \citep[e.g.,][]{Armitage96, Matt10, Matt12}. 
$J_{\star}$ is altered by external torques $\Gamma_\mathrm{ext}$ of which, following the formulation in \cite{Gallet19}, we model three torque components:

\textit{(1) Accretion torque, $\Gamma_\mathrm{acc}$:}
The accretion of disk material adds mass and angular momentum to the star.  $\Gamma_\mathrm{acc}$ is the product of the accretion rate onto the star and the specific angular momentum of the disk at the inner disk boundary \citep[e.g.,][]{Gallet19}{}{}.
This torque contribution is positive, which means that the star spins up due to the accretion of disk material.

\textit{(2) Accretion-powered stellar winds (APSW), $\Gamma_\mathrm{W}$:}
In the picture of a magnetized stellar outflow, the ejected material is magneto-centrifugally accelerated along stellar magnetic field lines.
As the wind propagates outward, it removes angular momentum from the star and results in stellar spin-down.
In this study, the driving force of this wind is considered to be accretion \citep[accretion-powered stellar wind, APSW, e.g.,][]{Matt05b, Gallet19}{}{} and the mass-loss rate scales as $\Dot{M}_\mathrm{W} = W \Dot{M}_\star$, with the APSW efficiency parameter, $W$.
The Alfven radius $r_\mathrm{A}$ acts as a lever arm for this stellar outflow and controls how much angular momentum is carried away \citep[e.g.,][]{Gallet19, Pantolmos20, Ireland21}{}{}.
We note that the order of $W$ as well as the driving forces of the APSW are still under debate \citep[see \sref{sec:apsw_W} or, e.g.,][]{Gallet19}{}{}.

\textit{(3) Magnetospheric ejections (ME), $\Gamma_\mathrm{ME}$:}
Angular momentum can be transferred between the star and the disk by magnetospheric ejections if the stellar magnetic field is strong enough to couple to the disk material \citep[e.g.,][]{Zanni13, Gallet13}{}{}.
Starting from vertical stellar magnetic field lines that co-rotate with the star, differential rotation between the star and the disk generates a toroidal field component.
The toroidal field results in a pressure component in the vertical direction that induces a periodic inflation and re-connection process of the magnetic field lines \citep[][]{Zanni13}{}{}.
The angular momentum transferred by this process, $\Gamma_\mathrm{ME}$, depends on the twist of the field lines, the stellar magnetic field strength, and the radial disk region that couples to the magnetic field lines \citep[e.g.,][]{Zanni13, Gallet19}{}{}.
The sign of $\Gamma_\mathrm{ME}$ results from the relation between the inner disk boundary, $r_\mathrm{trunc}$, and the co-rotation radius, $r_\mathrm{cor}$.
If the ratio between the truncation radius and the co-rotation radius is smaller than a given value $K_\mathrm{ME}$, $r_\mathrm{trunc}/r_\mathrm{cor} < K_\mathrm{ME}$, the MEs spin up the star.
For larger truncation radii, $r_\mathrm{trunc}/r_\mathrm{cor} \geq K_\mathrm{ME}$, the MEs spin down the star.
The dimensionless parameter, $K_\mathrm{ME}$, results from a best-fit value based on multi-dimensional models of the star-disk interaction \citep[e.g.,][]{Gallet19, Pantolmos20}{}{}.
The value of $K_\mathrm{ME}$ varies between different studies and ranges between 0.60 \citep[][]{Pantolmos20}{}{} and 0.79 \citep[][]{Gallet19}{}{}.
In this study, we choose the latter.

% The stellar angular momentum $J_{\star}$ is altered by external torques $\Gamma_\mathrm{ext}$ of which, following \cite{Gallet19}, we model three: (1) a positive torque $\Gamma_\mathrm{acc}$ from the accretion of disk material on the star at a rate $\Dot{M}_\mathrm{disk}$, causing spin-up; (2) a negative torque $\Gamma_\mathrm{W}$ from an accretion-powered stellar wind (APSW) with a mass loss rate $\Dot{M}_\mathrm{W}$ driven by a certain fraction $W$ of the disk's accretion rate ($\Dot{M}_\mathrm{W} = W \Dot{M}_\mathrm{disk}$); 
% and (3) torques arising from magnetospheric ejections (MEs) $\Gamma_\mathrm{ME}$ of either sign depending on the position of the inner disk radius $r_\mathrm{in}$ with respect to the co-rotation radius $r_\mathrm{cor}$ \citep[][]{Zanni13, Gallet13}.  The lever arm of the  APSW is taken to be the Alfv\'en radius of the wind $r_\mathrm{A}$. 
% The respective formulations of $\Gamma_\mathrm{acc}$, $\Gamma_\mathrm{W}$, $r_\mathrm{A}$ and $\Gamma_\mathrm{ME}$ are as described in \cite{Gallet19}. 

Internal re-distribution of angular momentum resulting from the development of a radiative core \citep[as described in][]{Gallet19} is neglected because very low-mass stars will remain fully convective during the disk lifetime and rotate as solid bodies. 
The evolution of the stellar rotation is governed by changes in AM and changes in the moment of inertia due to mass accretion and pre-main sequence contraction \citep[e.g.,][]{Matt10, Gallet19}. The rate of stellar mass change is calculated by the hydrodynamic disk model as the disk accretion rate at the inner boundary and we describe the stellar radius evolution by Kelvin-Helmholtz contraction with an index 3/2 polytrope \citep[e.g.,][]{Collier93, Matt10, Matt12}, resulting in
\begin{equation}\label{eq:stellar_radius_evolution}
    \Dot{R}_\star = 2 \frac{R_\star}{M_\star} \Dot{M}_\star - \frac{28 \pi \sigma R_\star^4 T_\mathrm{eff}^4}{3 \mathrm{G} M_\star^2} \, ,
\end{equation}
with the stellar effective temperature $T_\mathrm{eff}$ and the Stefan-Boltzmann constant $\sigma$. The deviation of \equ{eq:stellar_radius_evolution} from radii predicted by the pre-main sequence model of \cite{Baraffe15} is only $\sim1$\% over the first $10$~Myr.

\subsection{Parameter range}\label{sec:para_range}

We define a parameter range for our simulation that agrees reasonably with observations and theoretical estimates of stellar and disk parameters.

\subsubsection{Stellar rotation period of T~Tauri stars}
\begin{figure}
    \centering
         \resizebox{\hsize}{!}{\includegraphics{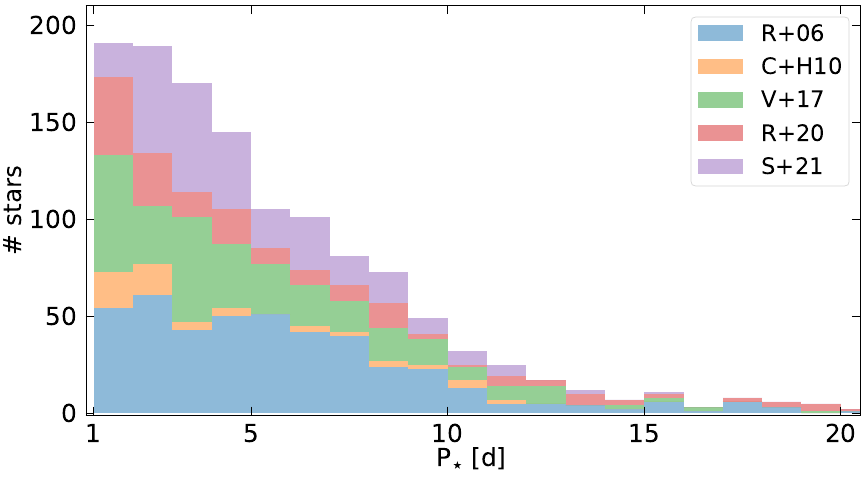}}
    \caption{
    Distribution of stellar rotation periods in young clusters: ONC \citep[R+06,][]{Rebull2006}, $\sigma$~Ori \citep[C+H10,][]{Cody2010}, NGC~2264 \citep[V+17,][]{Venuti2017}, Taurus \citep[R+20,][]{Rebull2020} and the Orion star forming region \citep[S+21,][]{Serna2021}.
    }
    \label{fig:onc_rot}
\end{figure}

Over the last two decades, the rotation periods of a large number of young ($\lesssim 10$~Myr) stars have been measured \citep[e.g.,][as well as other publications by the YSOVAR consortium]{herbst02, Rebull2006, Cody2010, Rebull2010, Rebull2015, Venuti2017, Rebull2020, Serna2021}.
In \fig{fig:onc_rot}, the distributions of rotation periods of five of these studies are shown. 
The majority of periods are located between $\sim 1$ and 10~days. 
The distribution decreases strongly towards $\sim 20$~days and there are very few stars with rotation periods $>20$~days.
This distribution of values, however, has to be treated with caution.
Studies based on TESS observations \citep[e.g.,][]{Rebull2015, Serna2021} are biased against periods longer than TESS’ lunar-synchronous orbit of 13.7~days \citep[e.g.,][]{Howard2021, Claytor2022} and longer undetected periods are possible\footnote{We refer to the sharp drop in rotation periods at $\sim 15$~days in Fig.~19 of \cite{Rebull2015}.}.
Furthermore, ground-based studies are subjected to the known effect of decreased sensitivity to longer periodic signals that are near the threshold of detection (only a few periods can be co-added for detection).
We adopt a conservative range of 1-10~days, recognizing that there are some stars with initial rotation periods outside this range.

\subsubsection{Stellar magnetic dipole field strength}

The dipole magnetic field strength is an important parameter for the spin model used in this study. 
Obtaining reliable estimates of $B_\star$, however, is challenging due to the limited number of observations and complex physical processes that generate the magnetic field \citep[e.g.,][]{Kochukhov21}.
There are different techniques for measuring magnetic fields \citep[e.g., the review of][and references therein]{Kochukhov21}. In the most common method, the broadening of magnetically-sensitive lines due to the Zeeman effect is measured and the strength of the magnetic field is described in terms of the Stokes~I parameter. 
We refer to this B-field measurement as $B_\mathrm{I}$. 
$B_\mathrm{I}$ is a measure of the total intensity of the surface magnetic field and it is not possible to differentiate between small and large-scale components. 
A more detailed technique, Zeeman-Doppler imaging (ZDI), resolves the distribution of line broadening on the stellar surface and obtains the Stokes~V parameter \citep[e.g.,][]{Johnstone14}. 
A large-scale magnetic field dipole component (which we refer to as $B_\mathrm{V}$) can be calculated from this. 
We note that there is the possibility that $B_\mathrm{V}$ can underestimate the strength of the magnetic field as it can be partially canceled out by higher-order components \citep[see the case of AU~Mic,][]{Kochukhov20}. If, in addition to the Stokes~I and V parameters, the Stokes~Q and U parameters are available, the overall structure of the stellar magnetic field and a more precise value for the large-scale field can be calculated \citep[e.g.,][]{Kochukhov20}. We refer to this large-scale field measurement as $B_\mathrm{QU}$.  Unfortunately, this requires observationally expensive high-resolution spectropolarimetry, and these have been performed on very few stars \citep[][]{Kochukhov20}.
The large-scale dipole field component (inferred indirectly based on Stokes~V or Stokes~Q and U parameters) usually dominates the magnetic star-disk interaction and governs the loss of AM due to the APSW \citep[e.g.,][]{Finnley18}; it is this component that is parameterized by the surface field parameter $B_\star$ in this study.

Aggravatingly, the stellar magnetic field evolves with time \citep[e.g.,][]{Folsom2016}. 
Young, fully convective stars have a strong, axisymmetric, and predominantly dipolar field.
As the stars evolve and develop a radiative core (stars with masses $\lesssim 0.3~\mathrm{M_\odot}$ stay fully convective during their entire lifetime), the field strength decreases, and the field complexity increases, reducing the dipole component \citep[e.g.,][]{Folsom2016}.
Based on the data for 8 T~Tauri stars, the large-scale field comprises up to $b_\mathrm{lsf} \approx 42$~\% of the total field intensity \citep[e.g.,][]{Lavail2019} ranging from $\sim 0.1$~kG up to $\sim 2$~kG in their sample.
For M~dwarfs, the total field strength can range up to $\sim 8$~kG \citep[][]{Reiners22}.
Additionally, we want to highlight the case of AU~Mic \citep[][]{Kochukhov20}.
AU~Mic is a 22~Myr-old, partially radiative M~dwarf with a total field intensity of 2.3~kG and a dipole component of 2.0~kG, corresponding to $b_\mathrm{lsf} \approx 87$~\% of the total intensity.
As AU~Mic has already developed a radiative core, the previously mentioned evolutionary trends \citep[][]{Folsom2016} allow the possibility of an even larger proportion of the large-scale field component $b_\mathrm{lsf}$ at younger ages.

Based on the aforementioned limitations and the small number of available observations, we note that reliable estimates of magnetic field strengths are difficult to obtain. 
The range of dipole field strengths from 0.1~kG to 2.0~kG used in this study covers most of the dipole field strengths inferred from observations. 
However, we do not exclude the possibility of higher and lower dipole field strengths.

\subsubsection{Initial disk accretion rates}

As shown in previous studies, the spin evolution of young stars depends on the initial accretion rate \citep[e.g.,][]{Gallet19}{}{}.
Therefore, it is important to define realistic initial accretion rates over the M~dwarf mass range at the starting point of our simulations, which is the beginning of the Class~II phase.
Age estimates of young Class~II objects start at $\gtrsim 1$~Myr but show a significant spread of $\sim1$~Myr, which complicates reliable estimates for the accretion rate at the beginning of the Class~II phase \citep[e.g.,][]{Manara2012, Fiorellino21, Testi2022}.
As a consequence, we motivate the choice of our initial accretion rates in this study not based on age estimates, but on the comparison of accretion rates of different evolutionary stages of the star-disk system.

\begin{figure}
    \centering
         \resizebox{\hsize}{!}{\includegraphics{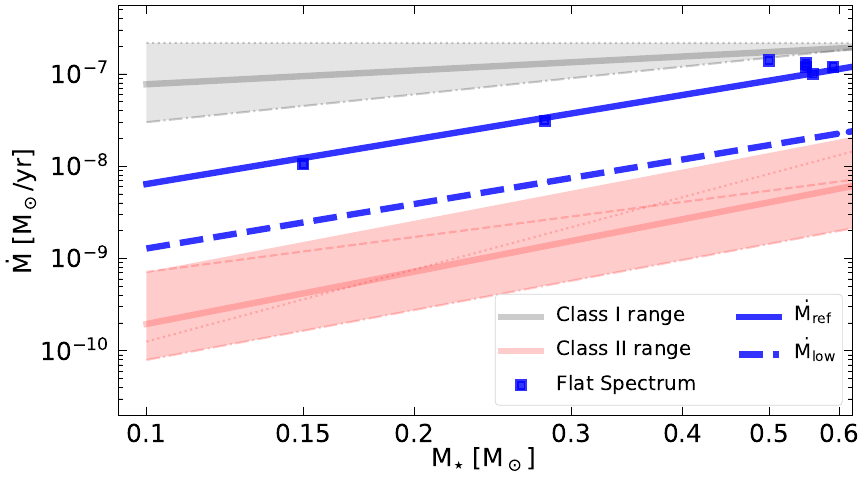}}
    \caption{
    Observed accretion rates of Class~I (grey), Class~II (red), and flat spectrum (blue) sources. The grey dash-dotted line is taken from \cite{Fiorellino21} and the grey dotted line from \cite{Caratti2012}.
    The gray-shaded area is limited by the two lines. The red dashed, dotted, and dash-dotted lines are taken from \cite{Manara2012}, \cite{Fiorellino21}, and \cite{Testi2022}, respectively. The red-shaded area has a minimum but constant logarithmic height so that all red lines are comprised. The thick grey and red lines indicate the mean values of the Class~I and Class~II objects. The blue squares represent accretion rates of flat spectrum sources (summarized in \tab{tab:flats}). The blue line is the best fit for the flat spectrum objects and the dashed line is five times less than that, just above the Class~II regime.
    }
    \label{fig:mdot_init}
\end{figure}

In \fig{fig:mdot_init}, observed accretion rates of Class~I \citep[assumed ages <1~Myr, e.g.,][]{Fiorellino2022}, Class~II, and ``flat spectrum" objects are summarized (see caption for references).
Flat spectrum (or transition) objects are characterized by a flat spectral energy distribution \citep[SED, e.g.,][]{Tobin2020}, indicating the transition between Class~I and Class~II.
Based on the observed sources, accretion rates of flat spectrum sources (blue squares) are located between Class~I (grey lines) and Class~II (red lines) objects.
Thus, we choose the flat spectrum sources as initial values for our simulations.
The solid blue line represents the fit and is taken as a reference value ($\Dot{M}_\mathrm{ref}$) for the initial accretion rates in the further course of this study and scales as
\begin{equation}
    \Dot{M}_\mathrm{init}(M_\star) = \Dot{M}_\mathrm{ref}(M_\star) = 3.7 \times 10^{-8} \left( \frac{M_\star}{0.3~M_\odot} \right)^{1.60}~\Msolpyr \, .
    \label{eq:mdot_init_flat}
\end{equation}
Due to large uncertainties in the observed accretion rate \citep[e.g.,][]{Caratti2012}, we also consider smaller (by a factor of 5) initial accretion rates $\Dot{M}_\mathrm{low}$; these values are located at the upper edge of the typical values for Class II objects for all values of $M_\star$ (dashed blue line in Fig. \ref{fig:mdot_init}).

%% file: sections/results.tex
\section{Results}
\label{sec:results}

We describe two sets of results: first, the influence of an accretion disk on stellar SAM and how the spin evolution of M dwarfs varies with different initial stellar and disk parameters (\sref{sec:critical_mdot}); and second, how the post-disk specific angular momentum (SAM) of M dwarfs depends on stellar and disk parameters (\sref{sec:j_final}). In all simulations, the disk is considered to be dissipated and simulations are stopped when the accretion rate falls below $10^{-13}$~$\mathrm{M_\odot~ yr}^{-1}$.   This is the minimum disk accretion rate that can be detected on M dwarfs with current observational methods \citep{Sicilia16}.  At this accretion rate, the residual disk typically has a mass of $\sim 10^{-6}~\mathrm{M_\odot}$ and ceases to influence the stellar rotation.

\begin{figure}
    \centering
         \resizebox{\hsize}{!}{\includegraphics{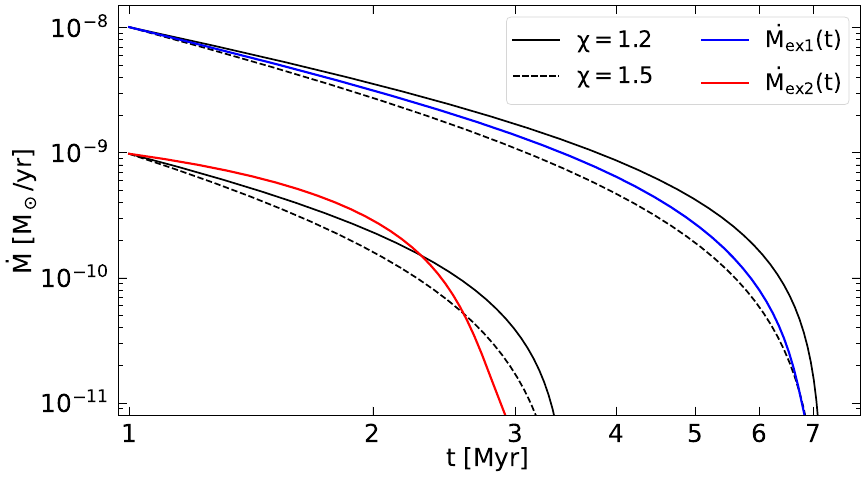}}
    \caption{
    Model disk accretion rates vs. time for two cases: The  $\Dot{M}_\mathrm{ex1}(t)$ (blue line) and $\Dot{M}_\mathrm{ex2}(t)$ (red line) cases are simulations of stars with $M_\star = 0.3~(0.1)~\mathrm{M_\odot}$ and a disk with a viscosity parameter $\alpha = 0.005~(0.010)$, respectively. The stellar magnetic field strength is the same ($B_\star = 1.0$~kG) in both cases. We compare these to empirical relations based on the power-law relation  $\Dot{M}\propto t^{-\chi}$ of \citet{Gallet19}) with disk lifetimes of 3.5 and 7.2 Myr and exponents $\chi = 1.2$ (solid black line) and $\chi = 1.5$ (dashed black line) \citep[][]{Vasconcelos2017, Gallet19}.
    }
    \label{fig:mdot_init_comp}
\end{figure}

Our simulations yield both disk accretion rates and lifetimes that are consistent with observations and empirical descriptions of disk accretion based on observations.  Figure \ref{fig:mdot_init_comp} compares disk accretion rate vs. time from our simulations to an empirical power-law relation $\Dot{M}\propto t^{-\chi}$ \citep[Eq.~3 in][]{Gallet19}, where $\chi$ is a fitting parameter.  
Two cases are shown: The blue (red) line $\Dot{M}_\mathrm{ex1}(t)$ ($\Dot{M}_\mathrm{ex2}(t)$) represents our simulations with a star with $M_\star = 0.3~(0.1)~\mathrm{M_\odot}$ and a disk with a viscous parameter $\alpha = 0.005~(0.010)$, respectively. The stellar magnetic field strength is the same ($B_\star = 1$~kG) in both cases.  
The power law exponent takes values of $\chi =1.2$ \citep[solid black line,][]{Gallet19} and $\chi = 1.5$ \citep[dashed black line,][]{Vasconcelos2017}. For disk lifetimes of 7.2 and 3.5~Myr, our simulations are in reasonable agreement with the empirical power law.
We note that the lifetimes among our simulations range between $\sim2$ and 11~Myr.
This range is comparable to disk lifetimes, inferred from observations \citep[e.g.,][and references therin]{Richert2018, Pfalzner2022}. 
Accretion rates that are initially $\gtrsim 10^{-8} M_{\odot}$~yr$^{-1}$ are also predicted to fall below $10^{-9} M_{\odot}$~yr$^{-1}$ within a few Myr (Fig. \ref{fig:mdot_init_comp}), consistent with the range of accretion rates observed in clusters at such ages \citep{Manara22}.

\subsection{The influence of a disk on stellar AM}\label{sec:critical_mdot}

To study the influence of an accretion disk on the central star, we consider different initial values for the stellar and disk AM and compare the final stellar AM to its initial value once the disk has dissipated. For a given initial stellar radius and mass, the stellar AM in our model can be changed by varying the initial rotational period $P_\mathrm{init}$.  Since the disk is in quasi-Keplerian rotation, its AM can be adjusted by changing the disk mass, which we do by varying its initial accretion rate $\Dot{M}_\mathrm{init}$ for a fixed $\alpha$ \citep[$\Dot{M}\propto \alpha \Sigma$ for a viscous accretion disk; see][]{Armitage2013}.

\begin{figure}
    \centering
         \resizebox{\hsize}{!}{\includegraphics{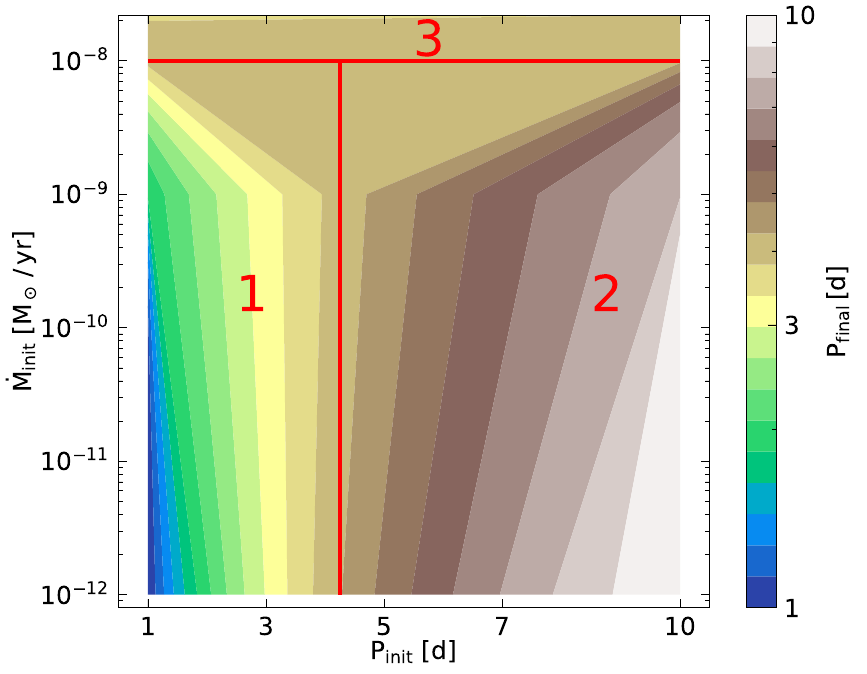}}
    \caption{Final stellar rotation period $P_\mathrm{final}$ for different initial periods $P_\mathrm{init}$ and disk accretion rates $\Dot{M}_\mathrm{init}$. The stellar and disk parameters are: $B_\star=1000G$, $M_\star=0.3 \mathrm{M_\odot}$, $\mathbf{W=2\%}$ and $\alpha=0.01$.
    We can divide the parameter space into three different regions, separated by red lines (see text).
    }
    \label{fig:contour}
\end{figure}

To illustrate the varying effect of disks with different initial accretion rates on the final stellar AM and rotation period, we consider, as a fiducial case,  a 0.3~$\mathrm{M_\odot}$ star with a photospheric magnetic field of $B_\star = 0.5$~kG.
The disk's viscous $\alpha$ parameter is set to $1\times10^{-2}$ and we set the efficiency of a stellar wind to $W=2~\%$. $P_\mathrm{init}$ is varied from 1 to 10 days and we vary $\Dot{M}_\mathrm{init}$ from $10^{-12}$ to $3\times 10^{-8}~\Msolpyr$ (see \sref{sec:para_range}).
Figure \ref{fig:contour} plots the final stellar rotation period at disk dissipation as a function of $P_\mathrm{init}$ and $\Dot{M}_\mathrm{init}$.  
For a more intuitive interpretation of this plot, we show the initial and final rotation periods instead of the SAM values. 
The same conclusions can be drawn for SAM because the duration of these simulations is short compared to the Kelvin-Helmholtz contraction time ($\tau_\mathrm{KH} \sim G M_\star^2 / (R_\star L_\star) $) of the star and thus the stellar moment of inertia does not change significantly.

In our stellar spin model (see \sref{sec:stellar_spin_model}), $P_\mathrm{final}$ in \fig{fig:contour} is influenced by accretion, the magnetic star-disk connection, and the APSW. 
These quantities are calculated self-consistently and depend on time and on stellar and disk parameters \citep[see][]{Steiner21}.
Contrary to the effects of accretion and the APSW, which (in our model) always spin up and spin down the star, respectively, MEs can result in a spin-up or spin-down torque that acts on the star.
As described in \sref{sec:stellar_spin_model} and, for example, in \cite{Gallet19}, the position between the truncation radius and the co-rotation radius determines the sign of this torque contribution.
MEs spin up the star if the truncation radius is located well within the co-rotation radius.
In case the truncation radius moves outward and toward the co-rotation radius, MEs can spin down the star \citep[e.g.,][]{Gallet19}{}{}.

For a given truncation radius ($r_\mathrm{trunc} < r_\mathrm{cor}$), the co-rotation radius of a star that spins up moves inward, and the distance between $r_\mathrm{trunc}$ and $r_\mathrm{cor}$ decreases.
Thus, the spin-up torque of the MEs is reduced or even turned into a spin-down torque in the case of a faster-rotating star.
Following the same logic, the MEs tend to increase the spin-up torque onto the star, if the star spins down.
For a given stellar rotation and co-rotation radius, an increased accretion rate can also alter the effects of the MEs.
A high accretion rate pushes the truncation radius inward and the distance between $r_\mathrm{trunc}$ and $r_\mathrm{cor}$ increases.
Thus, the MEs tend to spin up the star during phases of high accretion rates.
Analogously, the distance between the truncation and co-rotation radius decreases during phases of low accretion rates, and the MEs spin down the star.

Furthermore, the strength and sign of the MEs depend on the magnetic field strength.
A strong stellar magnetic field strength results in larger truncation radii.
As a consequence, the MEs tend to show a spin-down tendency for strong stellar magnetic field strength.
For weak stellar magnetic fields, on the other hand, the distance between the truncation and co-rotation radius increases, and the MEs show a spin-up tendency.
Furthermore, the magnitude of the torque due to MEs is proportional to $B_\star^2$ \citep[e.g.,][]{Gallet19}{}{} and the overall effect of MEs on the stellar spin evolution increases in case of strong magnetic field strengths.

We note that the sign and magnitude of the torque due to MEs, $\Gamma_\mathrm{ME}$, are strongly coupled to different (in turn coupled) parameters such as the accretion rate, stellar rotation period, and magnetic field strength.
We show the evolution of $\Gamma_\mathrm{ME}$ in \fig{fig:ME} for different configurations.
As expected, the MEs result in a spin-up torque for high accretion rates during the beginning of our simulations, weak magnetic field strengths, or slow rotation periods.
For fast-rotating stars and strong magnetic field strengths during phases of low accretion rates (especially toward the end of the disk lifetime), MEs spin down the star.
Compared to the spin-down torque of APSW (see \fig{fig:ME}), the effect of MEs is comparable for very fast-rotating stars. 
MEs are even the dominant source of stellar AM loss during the last $\gtrsim 1$~Myr of the disk lifetime.

We can divide the parameter space in \fig{fig:contour} into three regions (demarcated by red lines). 
In Region~1~(2), the star-disk interaction causes the stellar period to increase (decrease), respectively. 
The influence of the disk, however, is not strong enough to erase the imprint of the initial stellar spin values.
A rapidly rotating star remains rapidly rotating compared to a slow rotator and the dispersion in the final states is comparable to the spread of initial values.  
In this regime, at the bottom of \fig{fig:contour}, the contours are more vertical. 
In Region~3, the accretion disk is the dominant factor that controls the stellar spin evolution during the disk phase, the initial conditions are forgotten, and $P_\mathrm{final}$ converges towards a narrower range of values, nearly independent of $P_\mathrm{init}$.
 This is the regime at the top of \fig{fig:contour} where the contours are nearly horizontal.  The star's initial rotation state is ``erased" by the interaction with the disk.  This phenomenon has been seen in previous studies \citep[e.g.,][]{Armitage96, Matt12} but the dependence of this behavior on star and disk parameters has not been investigated.

To investigate how this behavior depends on stellar and disk properties other than accretion rate, we vary these parameters: the strength of the star's magnetic dipole field at its surface $B_\star \in [0.5,1\,\mathrm{kG}]$ (see \sref{sec:para_range}), the initial stellar mass $M_\star \in [0.3,0.6\,\mathrm{M_\odot}]$, and the disk viscous parameter $\alpha \in [0.005, 0.01]$.
The resulting effect of the accretion disk on the stellar spin evolution is visualized by comparing the ratios between maximum and minimum values of the stellar rotational period and the specific angular momentum after the disk has been dispersed relative to the ratios of the initial values. 
A parameter $C$ is introduced to quantify the change of these ratios by the following relations
\begin{equation}\label{eq:c_p}
    \left( \frac{P_\mathrm{max}}{P_\mathrm{min}} \right)_\mathrm{final} = \left( \frac{P_\mathrm{max}}{P_\mathrm{min}} \right)^{C_\mathrm{P}}_\mathrm{init}
\end{equation}
and
\begin{equation}\label{eq:c_j}
    \left( \frac{j_\mathrm{max}}{j_\mathrm{min}} \right)_\mathrm{final} = \left( \frac{j_\mathrm{max}}{j_\mathrm{min}} \right)^{C_\mathrm{j}}_\mathrm{init} \, .
\end{equation}
The left-hand sides of \equs{eq:c_p}{eq:c_j} denote the stellar rotational periods and specific angular momentum values after the disk has been dissolved (final values) and the right-hand sides the respective initial values. $C$ can be understood as the slope of the contours in a log-log plot (\fig{fig:contour}).
A value of $C$ close to $1$ corresponds to a weak influence of the accretion disk. A strong influence on the stellar spin evolution on the other hand is indicated by $C\ll1$. In this case, the initial rotational values are "forgotten" and the star's final state is independent of the initial rotation.  We consider $C = 0.01$ (corresponding to a final difference of $\approx 2\%$) sufficiently low for final conditions to be convergent and define the initial accretion rate at which this condition is reached as a critical accretion rate $\Dot{M}_\mathrm{crit}$.

\begin{figure*}[ht!]
    \centering
         \resizebox{\hsize}{!}{\includegraphics{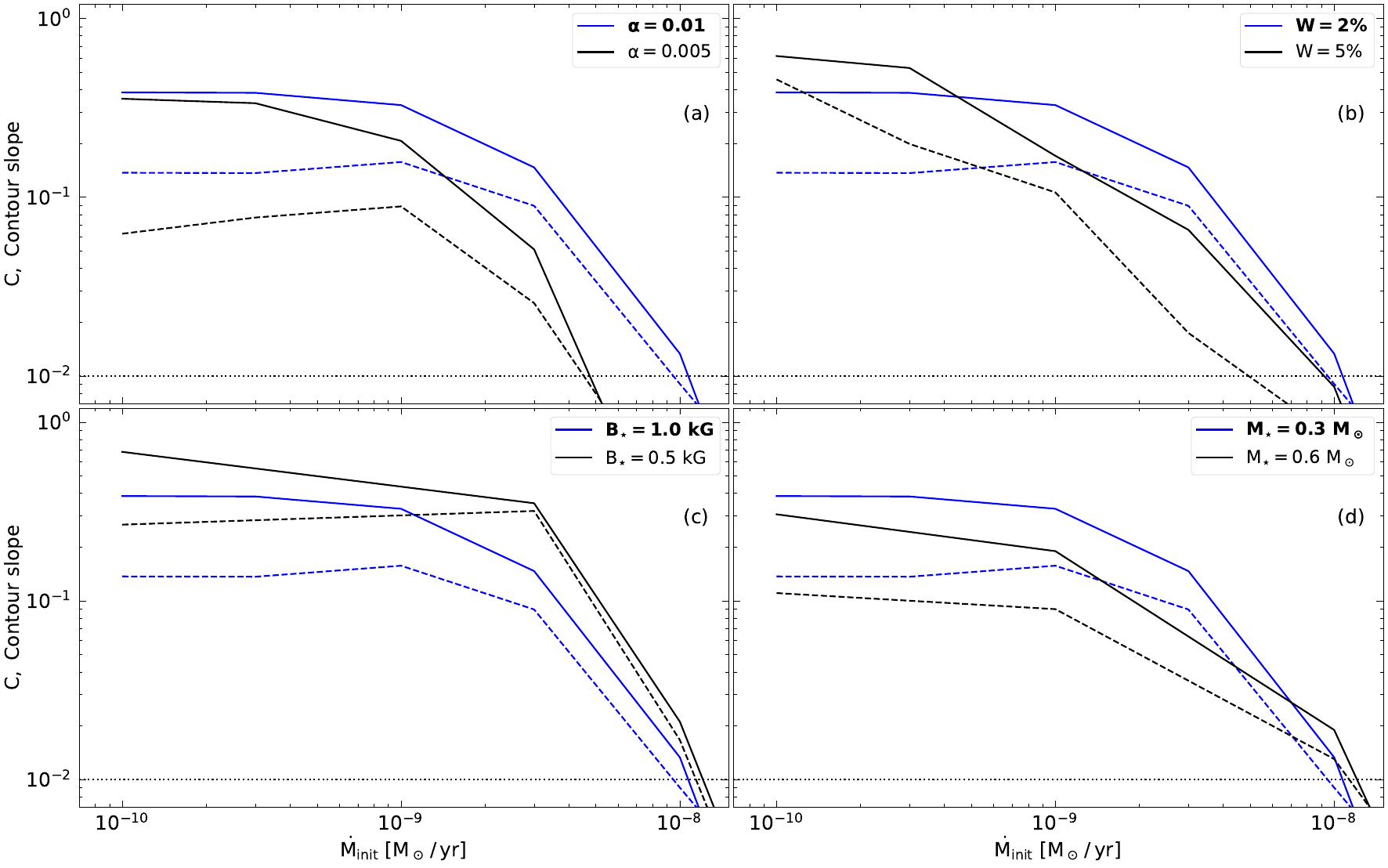}}
    \caption{
    The $C$-parameter (the gradient of the contours in \fig{fig:contour}) with respect to initial disk accretion rates $\Dot{M}_\mathrm{init}$. The solid lines represent $C_\mathrm{P}$ and the dashed lines $C_\mathrm{j}$ values, respectively. Each subplot shows variations in (a) viscous $\alpha$ parameter: 0.01 (blue lines) and 0.005 (black lines). The stellar mass is 0.3~$\mathrm{M_\odot}$ with a magnetic dipole field strength of $B_\star = 1$~kG. (b) $W$ values: 2\% (blue lines) and 5\% (black lines). The stellar mass is 0.3~$\mathrm{M_\odot}$ with a magnetic dipole field strength of $B_\star = 1$~kG and a viscosity parameter of $\alpha = 0.01$. (c) $B_\star$ values: 1.0~kG (blue lines) and 0.5~kG (black lines). The stellar mass is 0.3~$\mathrm{M_\odot}$ with a viscosity parameter of $\alpha = 0.01$. (d)  $M_\star$ values: 0.3~$\mathrm{M_\odot}$ (blue lines) and 0.6~$\mathrm{M_\odot}$ (black lines). The viscosity parameter is $\alpha = 0.01$. The threshold of $C = 0.01$ for the star to be in the "locked"-state is indicated by the horizontal dotted line. For better visualization, the highlighted parameters show the reference value in the other subplots.
    }
    \label{fig:C_all}
\end{figure*}

We quantify the effect of variation of the parameters disk viscosity ($\alpha$), stellar wind ($W$), stellar magnetic field ($B_\star$), and stellar mass ($M_\star$) on  $\Dot{M}_\mathrm{crit}$ in \fig{fig:C_all}. 
These parameters, $\alpha$ in Panel a, $W$ in Panel b,  $B_\star$ in Panel c, and  $M_\star$ in Panel d, determine the spin-down behavior.
Therefore, for each set of parameters, we can identify a value of $\Dot{M}_\mathrm{crit}$ that is $\sim 10^{-8}~\Msolpyr$.
Above this value, the stellar rotational period, as well as the SAM, do not depend on their initial values.

\subsection{
Relations for the post-disk specific angular momentum
}\label{sec:j_final}

We examine the dependencies of the SAM after the disk phase on stellar and disk parameters.
We compare these to the distribution of SAM among M dwarfs in the Upper Scorpius star-forming region reported by \cite{Somers17}.  
This sample is especially suitable for studying the AM distribution after the disk phase because the cluster has an age of $\sim10$~Myr \citep[][]{Feiden16} and most accretion disks dissipate by that time \citep[][]{Mamajek09}; also there's been no significant loss of AM from a thermal stellar wind up to this age.
\citet{Somers17} provided a mean and standard deviation for SAM in mass bins of $0.10 - 0.25~\mathrm{M_\odot}$, $0.25 - 0.40~\mathrm{M_\odot}$ and $0.40 - 0.60~\mathrm{M_\odot}$, respectively.

\subsubsection{Influence of stellar parameters on the post-disk SAM}\label{sec:reference_case}

\begin{figure*}[ht!]
    \centering
         \resizebox{\hsize}{!}{\includegraphics{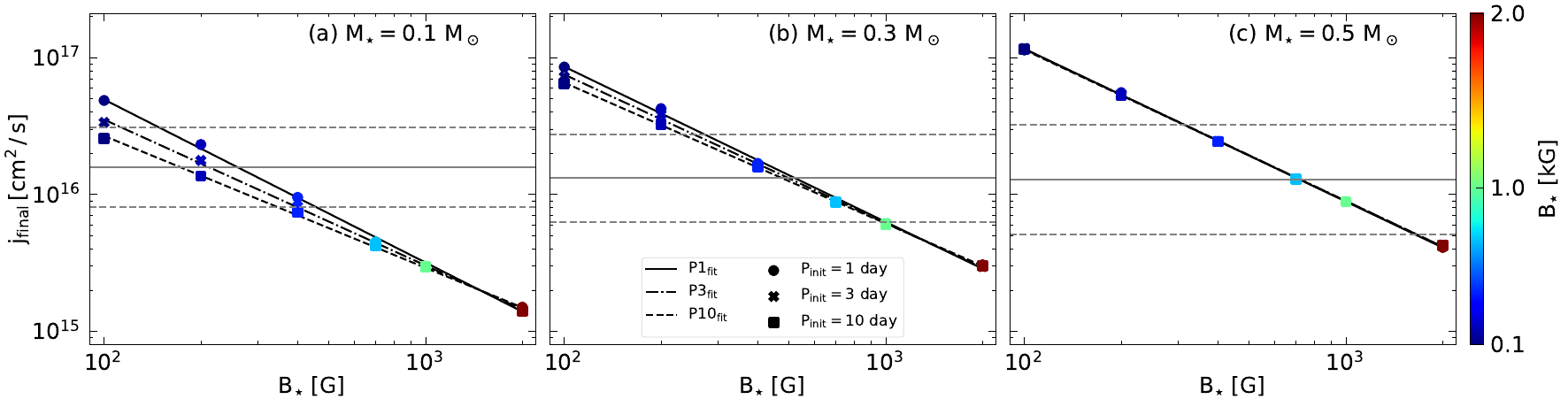}}
    \caption{
    The SAM after the disk phase $j_\mathrm{final}$ for different stellar magnetic field values and stellar masses of 0.1~$M_\odot$ (Panel a), 0.3~$M_\odot$ (Panel b), and 0.5~$M_\odot$ (Panel c).
    The colored circles, crosses, and squares represent initial rotation periods of 1, 3, and 10~days, respectively, and the solid, dash-dotted, and dashed lines show the respective fits.
    Mean and standard deviation values for the SAM taken from \cite{Somers17} are represented by grey solid and dashed lines, respectively.
    }
    \label{fig:B_high}
\end{figure*}

We perform a set of simulations to examine the effect of varying stellar mass and magnetic field on the final (post-disk) SAM of the star, $j_\mathrm{final}$. The initial stellar mass is set to 0.1, 0.3, 0.5~$\mathrm{M_\odot}$, reflecting the three mass bins of \cite{Somers17}. 
We note that during the simulations the stellar mass is increased by $\lesssim 5$~\% due to the accretion of disk material.
Three different initial stellar rotation periods are assumed: 1, 3, and 10~days. 
We adopt initial accretion rates that correspond to an evolutionary stage just preceding the classical T Tau stage, namely those of flat-spectrum transitional Class I to Class II sources \citep{Tobin2020}; these empirically determined accretion rates are mass dependent (see \equo{eq:mdot_init_flat}).  
In addition, the magnetic dipole field strength is set to values of 0.1, 0.2, 0.4, 0.7, 1.0, and 2.0~kG. 
The viscous parameter is set to $\alpha= 0.01$.
The SAM after the disk phase $j_\mathrm{final}$ for each possible combination is shown in \fig{fig:B_high}. 
The colored circles, crosses, and squares represent initial rotation periods of 1, 3, and 10~days, respectively, and the solid, dash-dotted, and dashed lines show the respective fits.
Mean and standard deviation values for the \textit{observed} SAM taken from \cite{Somers17} for each mass bin are represented by grey solid and dashed lines, respectively.

For stellar masses, $\gtrsim 0.3~\mathrm{M_\odot}$ the initial accretion rate is $\gtrsim \Dot{M}_\mathrm{crit}$ (see \equ{eq:mdot_init_flat}) and rotation period at the end of the disk phase is largely independent of the initial period (see Panel c).
For stellar masses, $< 0.3~\mathrm{M_\odot}$, the initial accretion rates are below the critical value and the initial rotation period influences the final value for $B_\star < 1$~kG (see panel a).
The ensemble of outcomes is well represented by power-laws of the form $j_\mathrm{final}\sim B_\star^{-b}$.
The magnetic field strengths of the intersection between the fitted data and the mean values from \cite{Somers17} ($\overline{B}_\star$) and the exponents $b$ are summarized in \tab{tab:Bhighs}.

\subsubsection{Influence of disk parameters on $j_\mathrm{final}$}

\begin{figure*}[ht!]
    \centering
         \resizebox{\hsize}{!}{\includegraphics{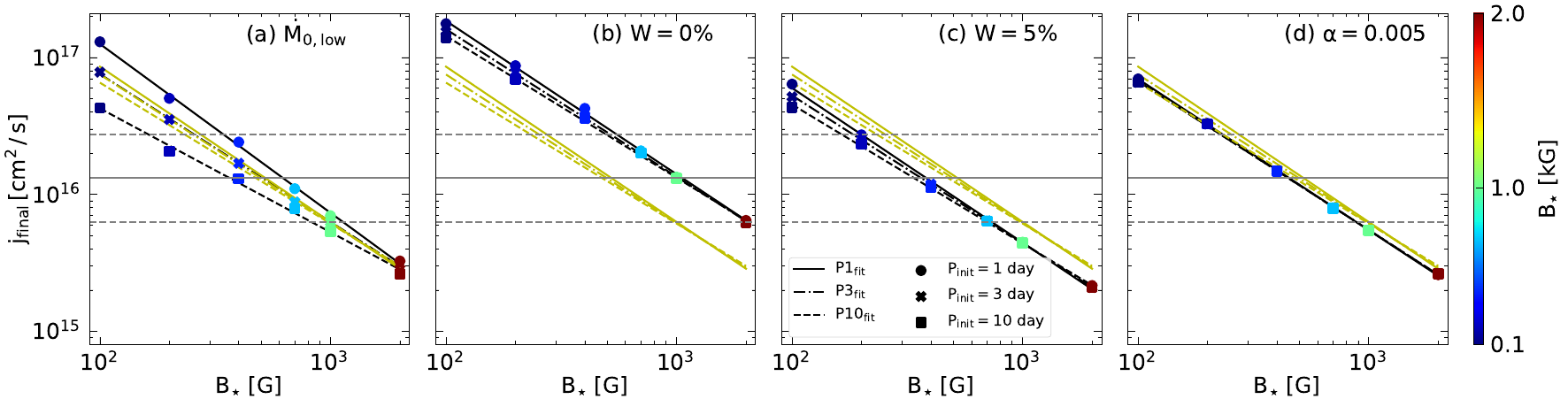}}
    \caption{
    The SAM after the disk phase $j_\mathrm{final}$ as a function of stellar magnetic field values (x-axis and colors) and for different initial rotation periods (different sets of point styles and curves). 
    Each panel shows the variation of one parameter with respect to the reference case in \fig{fig:B_high}.
    In Panel (a), $\Dot{M}_\mathrm{low}$ is used as the initial accretion rate.
    In Panel (b) and (c), the APSW efficiency is set to $0~\%$ and $5~\%$, respectively.
    In Panel (d), the viscous $\alpha$ parameter is set to 0.005.
    In all cases, the stellar mass is $0.3~\mathrm{M_\odot}$.
    The symbols and lines are identical to \fig{fig:B_high}.
    For comparison, the yellow lines show the fits from \fig{fig:B_high} for a stellar mass of 0.3~$\mathrm{M\odot}$
    }
    \label{fig:B_vary}
\end{figure*}

To study how the $j_\mathrm{final}$ relations derived in \sref{sec:reference_case} depend on the assumed values for disk parameters, we vary the initial accretion rate, the viscosity value $\alpha$, as well as the APSW efficiency $W$ and compare the results to the standard cases described in the previous subsection.
In \fig{fig:B_vary}, the results are shown and in \tab{tab:B_vary}, the magnetic field strengths of the intersection between the fitted data and the mean values from \cite{Somers17} ($\overline{B}_\star$) and the exponents $b$ are reported.
In the following cases, the initial stellar mass is set to $0.3~\mathrm{M_\odot}$.
For comparison, the yellow lines show the fits from \fig{fig:B_high} for a stellar mass of 0.3~$\mathrm{M\odot}$

We reduce the initial accretion rate by a factor of 5 compared to the values generated by \equ{eq:mdot_init_flat}.
At the lower initial accretion rates $\Dot{M}_\mathrm{low}$ (Panel a), the star-disk interaction is weaker than in the reference case, and $j_\mathrm{final}$ is more strongly influenced by the initial conditions. 
As a result, the difference between $\overline{B}_\star$ and $b$ for fast and slow initial rotation periods is larger.
More intuitively, if the predicted values of $j_\mathrm{final}$ are to agree with the observations, a larger variation in $B_\star$ and the exponent $b$ is needed to compensate for variations in the initial rotation period.

The amount of AM that is removed by the APSW is controlled by the efficiency parameter $W$.
As a consequence, the values for $j_\mathrm{final}$ are shifted in the vertical direction towards higher (lower) values for smaller (larger) values of $W$, respectively.
The values for the magnetic field strengths of the intersection between the fitted data and the mean values from \cite{Somers17}, $\overline{B}_\star$, are nearly equally increased (reduced) by 530~G (137~G) in case of $W=0\%$ ($W=5\%$) (see Panels b and c in \fig{fig:B_vary} and \tab{tab:B_vary}).
At a given accretion rate, a lower value of $\alpha$ increases the disk mass and lifetime, allowing star-disk interaction to influence the stellar AM over a longer time period and reducing the critical accretion rate (see Panel a in \fig{fig:C_all}).
The values in $\overline{B}_\star$ and $b$ for different $P_\mathrm{init}$ values converge and the initial conditions are now (nearly) completely forgotten (Panel d in \fig{fig:B_vary}).

The stellar SAM after the disk phase can be described by a power law such as
\begin{equation}\label{eq:jfinal}
    j_\mathrm{final} = j_\mathrm{ref} \left(\frac{B_\star}{0.5~\mathrm{kG}} \right)^{-b} \, ,
\end{equation}
with the proportionality factor $j_\mathrm{ref}$, depending on the stellar mass and, if  $\Dot{M}_\mathrm{init} < \Dot{M}_\mathrm{crit}$, also on the initial stellar rotation period $P_\mathrm{init}$. 
The values for $j_\mathrm{ref}$ for our reference case are summarized in \tab{tab:Bhighs}.

\begin{table}[ht]

\centering
\caption{
Values of $\overline{B}_\star$, $b$, and $j_\mathrm{ref}$ for models with $\Dot{M}_\mathrm{ref}$. The units for $M_\star$, $P_\mathrm{init}$, $\overline{B}_\star$, and $j_\mathrm{ref}$ are $\mathrm{M_\odot}$, days, G, and $10^{16}~\mathrm{cm^2/s}$ respectively.
}
\begin{tabular}{c | c c c }         
\hline\hline                       
$M_\star$ & 0.1 & 0.3 & 0.5\\
$P_\mathrm{init}$ & 1,3,10 & 1,3,10 & 1,3,10 \\
\hline  

$\overline{B}_\star$ & 259,211,172 & 521,497,475 & 720,718,715 \\ 
$b$ & 1.19,1.06,0.96 & 1.13,1.08,1.02 & 1.11,1.11,1.10\\
$j_\mathrm{ref}$ & 0.73,0.63,0.57 & 1.38,1.31,1.25 & 1.93,1.92,1.92\\

\hline       
\end{tabular}

\label{tab:Bhighs}  
\end{table}

\begin{table*}[ht]

\centering
\caption{
Values of $\overline{B}_\star$ and $b$ for models with $\Dot{M}_\mathrm{low}$, $W=5~\%$, and $\alpha = 0.005$. The values $\Delta_\mathrm{\overline{B}_\star}$ and $\Delta_\mathrm{b}$ show the difference with respect to the models with $\Dot{M}_\mathrm{ref}$. The initial stellar mass is 0.3~$\mathrm{M_\odot}$ for all models. The units for $P_\mathrm{init}$, $\overline{B}_\star$, and $j_\mathrm{ref}$ are days, G, and $10^{16}~\mathrm{cm^2/s}$, respectively.
}              
\begin{tabular}{c | c c c c }         
\hline\hline                       
 & $\Dot{M}_\mathrm{low}$ & $W=0~\%$ & $W=5~\%$ & $\alpha = 0.005$\\
$P_\mathrm{init}$ & 1,3,10 & 1,3,10 & 1,3,10 & 1,3,10 \\
\hline  

$\overline{B}_\star$ & 536, 426, 298 & 1058, 1021, 1005 & 381, 362, 338 & 461, 458, 456 \\ 
$\Delta_\mathrm{\overline{B}_\star}$ & +15, -71, -177 & +537,+524,+530 & -140,-135,-137 & -60,-39,-19 \\
$b$ & 1.23, 1.08, 0.90 & 1.12, 1.08, 1.03 & 1.13, 1.08, 1.01 & 1.10, 1.09, 1.09\\
$\Delta_\mathrm{b}$ & +0.10,<0.01,-0.12 & -0.01,<0.01,+0.01 & <0.01,<0.01,-0.01 & -0.03,+0.01,+0.07 \\
% $j_\mathrm{ref}$ & & & &\\
% $\Delta_\mathrm{j_\mathrm{ref}}$ & & & &\\

\hline       
\end{tabular}

\label{tab:B_vary}  
\end{table*}

%% file: sections/conclusion.tex
\section{Discussion}
\label{sec:conclusion}

% - Summarize what was done in this paper

We now wish to discuss the consistency of our results with observations, resulting implications, limitations, and uncertainties due to initial conditions and the adopted ranges of our principal parameters.

\subsection{Comparison of $\overline{B}_\star$ with inferred dipole field strengths}

\begin{figure}
    \centering
         \resizebox{\hsize}{!}{\includegraphics{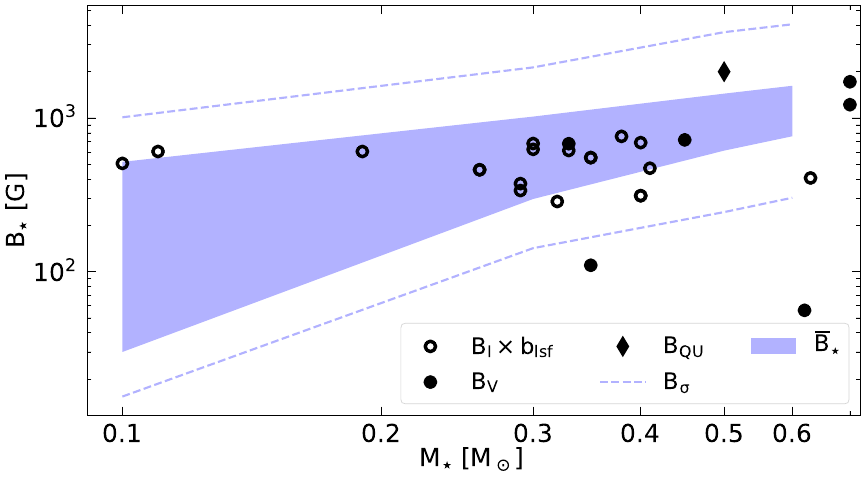}}
    \caption{
    Observed dipole field strengths of young stars with respect to stellar mass. Solid black circles show dipole field strengths inferred by the Stokes~V parameter ($B_\mathrm{V}$), open circles indicate an estimate based on the total field intensity ($B_\mathrm{I}$) and a dipole field strength fraction of $b_\mathrm{lsf}=22~\%$ \citep[][]{Lavail2019}, and the black diamond indicates the dipole field strength of AU~Mic, the only young star in our sample with direct measurement of $B_\mathrm{QU}$.
    The errors are expected to reach up to one order of magnitude.
    The range of the required magnetic field strengths to match the observed median value of the SAM distribution \citep[][]{Somers17} is marked in blue ($\overline{B}_\star$) and the respective standard deviations given in \cite{Somers17} are indicated by the blue dashed lines ($B_\mathrm{\sigma}$).
    }
    \label{fig:Bcomp}
\end{figure}

We compare the values of $\overline{B}_\star$ that required to explain the observed SAM to the dipole field strengths observed or inferred on young, low-mass stars.  In \fig{fig:Bcomp}, dipole magnetic field strengths are shown with respect to the stellar mass. 
Solid black circles show dipole field strengths inferred by the Stokes~V parameter ($B_\mathrm{V}$), open circles indicate an estimate based on the total field intensity ($B_\mathrm{I}$) and a dipole field strength fraction of $b_\mathrm{lsf}=22~\%$ \citep[][]{Lavail2019}, and the black diamond indicates the dipole field strength of AU~Mic, the only young star in our sample with direct measurement of $B_\mathrm{QU}$.
The data is taken from \cite{Yang2011}, \cite{Johnstone14}, and \tab{tab:obs_stars}.
It is important to note that, for this comparison, young accreting and non-accreting stars are combined to increase the number of available data points. 
To minimize the effect of an evolving magnetic field, the age of the non-accreting stars is restricted to the disk dispersal age (see \tab{tab:obs_stars}).

The range of the required magnetic field strengths to match the observed median value of the SAM distribution \citep[][]{Somers17} is marked in blue ($\overline{B}_\star$) and the respective standard deviations given in \cite{Somers17} are indicated by the blue dashed lines ($B_\mathrm{\sigma}$).
The changes of $\overline{B}_\star$ due to different disk and stellar parameters ($\Delta_\mathrm{\overline{B}_\star}$, see \tab{tab:B_vary}) are extrapolated from 0.3~$M_\odot$ over the M~dwarf mass range.
To reproduce the observed mass-indepdent distrubiton of SAM, our model requires a positive relation between stellar mass and magnetic field strength.
The observed values can explain the disk-star coupling required to explain the post-disk SAM distribution \citep[][]{Somers17}.
The data, however, are not sufficient and the uncertainties are too large to test the required mass dependence.

Theoretical justification for a mass-dependent magnetic field strength is described by \cite{Browning16}.
By equating the magnetic energy density $B_\star^2$ to the kinetic energy density in the turbulent flow of the star's convective envelope $\rho v_{\rm c}^2$, with $v_{\rm c}$ the convective velocity, \citet{Browning16} derive a theoretical kinetic limit to the magnetic field strength on rapidly-rotating stars.
Since turbulent convection is responsible for energy transport from the stellar interior, $B_\star$ can then be related to the luminosity of the star and thus to $T_{\rm eff}$ via the Stefan-Boltzmann relation: $B_\star \propto \left(M_\star T_{\rm eff}^8 / R_\star^3\right)^{1/6}$. Power-law fits of $T_{\rm eff}$ vs. $M_\star$ and $R_\star$ vs. $M_\star$ for \citet{Baraffe15} $\le$10 Myr isochrones return power-law indices of $\approx$0.15 and 0.33, respectively, depending on age.  
This yields $B_\star \propto M_\star^{0.2}$.
If Ohmic dissipation and magnetic buoyancy are considered in addition \citep[][]{Browning16}, the magnetic field strength scales like $B_\star \propto M_\star^{0.3}$.
Compared to our results, this relation shows a slightly weaker trend in the same direction, but can at least partially explain the observed distribution of SAM.

\subsection{Comparisons with individual stars after the disk phase}\label{sec:con_obs}

We compare our results for $j_\mathrm{final}$ to available measurements of stellar rotation, mass, and magnetic field strength for stars that are old enough for disks to have dissipated, but not significantly older than $\sim 100$~Myr so that the AM lost in a magnetized wind is small.
Unfortunately, the set of published observations that are relevant to our case is very limited. 
Due to the nature of the different measurement techniques, we use $B_\mathrm{I}$ as an upper limit and $B_\mathrm{V}$ as a lower limit of the surface field associated with the large-scale dipole component. 
While there are $\sim 50$ stars that have estimates of $B_\mathrm{I}$ and $B_\mathrm{V}$ \citep[][]{Kochukhov21}, only five of these fall within the required mass ($<0.6~\mathrm{M_\odot}$) and age ranges ($\lesssim 100$~Myr) and $B_\mathrm{QU}$ is only available for AU~Mic \citep[][]{Kochukhov20}. In total, 12 stars are chosen to be shown in this comparison (see \tab{tab:obs_stars}). For some stars, there is no large-scale magnetic field measurement available.  Finally, large uncertainties in the observed quantities impede the comparison.   Our analysis should be considered as a coarse interpretation of possible trends and an order of magnitude consideration.

\begin{figure}
    \centering
         \resizebox{\hsize}{!}{\includegraphics{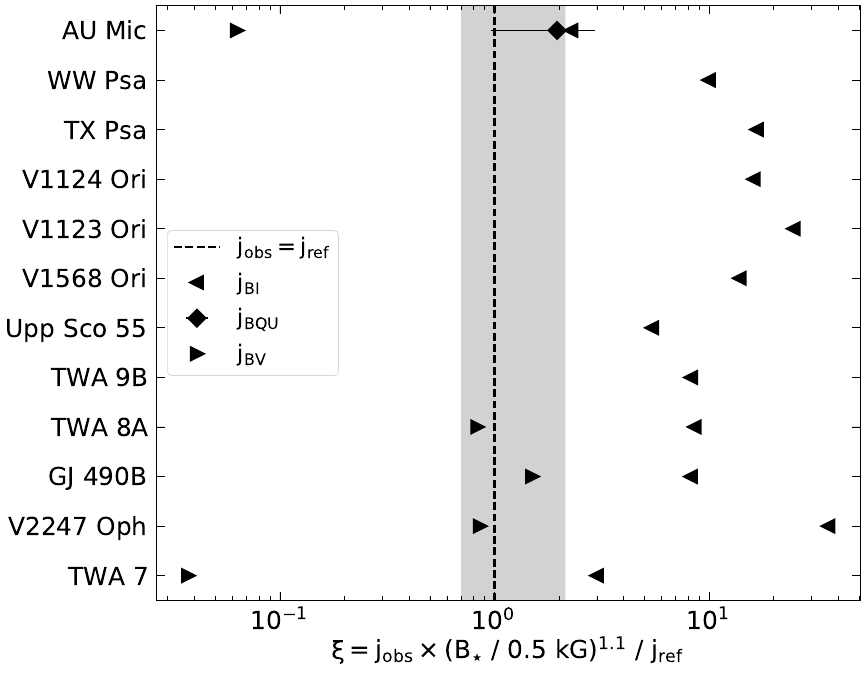}}
    \caption{
    The ratio of the observed to predicted SAM $j_\mathrm{obs}/j_\mathrm{final}$. To facilitate the comparison, $j_\mathrm{obs}$ is multiplied by a factor of $(B_\star/\mathrm{0.5~kG})^{1.1}$ (see text). 
    The vertical dashed line indicates a perfect agreement between the observed value and our model $j_\mathrm{ref}(\mathrm{M_\star})$.
    The grey area accounts for different stellar and disk parameters.
    The SAM values $j_\mathrm{BI}$ and $j_\mathrm{BV}$ are based on $B_\mathrm{I}$ and $B_\mathrm{V}$ functions as the upper and lower boundary for the stellar SAM.
    They are indicated by left and right-facing triangles, respectively. The single measured value (and uncertainties) of $j_\mathrm{BQU}$ (diamond marker) is based on $B_\mathrm{QU}$ for the star AU Mic. The measurements are tabulated in Table \ref{tab:obs_stars}.
    }
    \label{fig:obs_after}
\end{figure}

The observations are summarized in \tab{tab:obs_stars} and we compare these to our predicted values of $j_\mathrm{ref}(M_\star)$ scaled by $B_\star^{1.1}$ (see \sref{sec:j_final}) in \fig{fig:obs_after}. 
The upper and lower limits of $\xi = j_\mathrm{obs}\times (B_\star/0.5~kG)^{1.1} / j_\mathrm{ref}$ due to $B_\mathrm{I}$ and $B_\mathrm{V}$ are indicated by left and right facing triangles, respectively. The more precise estimate for AU~Mic based on $B_\mathrm{QU}$ (black diamond) is plotted with an error bar.  
A value of $\xi = 1$ (vertical dashed line) indicates a perfect agreement with our reference case. 
Due to the variations in stellar and disk parameters, $j_\mathrm{ref}$ can vary (see \fig{fig:B_vary}), which is indicated by the grey area.
This grey area falls within the range bracketed by the observed values of $B_\mathrm{I}$ and $B_\mathrm{V}$.  
Note that we assume $P_\mathrm{init} = 3$~days for all stellar masses for this comparison.
The agreement is encouraging but the limited number of observations, all but one of which are lower or upper limits on the magnetic field, does not allow us to draw firm conclusions.
Many more detailed measurements of the magnetic fields of young M dwarfs are needed for a more robust test of our models.

\subsection{Accretion-powered stellar wind (APSW) efficiency $W$}\label{sec:apsw_W}

Winds, in this study represented by APSW, can strongly affect the distribution of stellar SAM as well as the structure and long-term evolution of its disk \citep[e.g.,][]{Matt05b, Gallet19}. APSW is a potential explanation for the existence of slowly-spinning stars, but their exact mechanisms and strength are the subject of active research \citep[see the discussion in][]{Gallet19}. Observations of Doppler-shifted line absorption features \citep[][]{Dupree05, Edwards06} as well as statistical evaluations of the observed ratio of outflow mass loss to disc accretion rate \citep[e.g.,][]{Cabrit02, Ferreira06} indicate that there are indeed strong outflows from young star-disk systems.
These outflows are expected to be too cool to originate from a thermal pressure gradient at the stellar corona where temperatures are $\sim 10^6$~K \citep[e.g.,][]{Gallet19} and their exact launch point and driving mechanism are not known \citep[e.g.,][]{Ireland21} but they are supposed to be disk winds launched in the inner disk, possibly by the magneto-centrifugal process \citep[e.g.,][]{Konigl11}; some of these winds can even form jets.
A possible alternative driving mechanism for winds is proposed in \cite{Decampli81} and \cite{Matt05b}. The accreting disk material can excite Alfvén waves originating at the stellar surface, traveling outwards and removing angular momentum from the star. 
The mass loss rate through these winds is expected to be $W \sim 2$~\% \citep[][]{Crammer08, Pantolmos20}.

While previous authors need large APSW values of $\sim 10\%$ to explain slowly spinning stars, which contradicts the results of \cite{Crammer08} and \cite{Pantolmos20}, we can reduce this value considerably to reproduce the observed rotational period of slowly rotating stars. 
In this work, we can limit $W$ to $\lesssim 5\%$, which is in better agreement with observational constraints and consistent with the theoretical predictions of \cite{Crammer08} and \cite{Pantolmos20}. 
The reduction of $W$ might be explained by the combined stellar and disk model that includes the back-reactions of the rotational evolution on the accretion disk and the spin-down effect of MEs, especially during phases of fast stellar rotation and towards the end of the disk lifetime.
Apart from APSW and MEs, other wind mechanisms can be important for star and disk evolution.
Especially, disk winds noticeably affect the disk evolution \citep[e.g.,][]{Konigl11}. By removing mass and (if magnetized) angular momentum from the disk before it can be accreted onto the star, the evolution of stellar spin evolution is affected. Likewise, photo-evaporation (internal or external) can influence the disk's lifetime and thus, the stellar spin evolution \citep[e.g.,][]{Roquette21}.
These effects will be added to a future version of our model.

\subsection{Comparing our model to large-scale parametric studies}

Our model predictions can be directly integrated into parametric-based modeling of the stellar spin evolution of young stars, including the influence of an accretion disk \citep[e.g.,][]{Vasconcelos2017}. 
While \cite{Vasconcelos2017} have conducted a large number of simulations ($\sim 2\times 10^{5}$), exceeding the number of our simulations by over two orders of magnitude, their model is parametric and the evolution of the accretion disk and the star-disk interaction is based on simplified assumptions.
The disk accretion rate is taken to be independent of stellar mass and the disk lifetime is estimated based on a free parameter.
During the disk lifetime, disk-locking is assumed (the stellar rotational period is constant over time) and important aspects like the influence of a stellar magnetic field or an accretion-powered stellar wind are ignored. 

\cite{Vasconcelos2017} can present useful guidelines regarding the rotational evolution of low-mass stars.
However, none of their models can reproduce all the available observational aspects (e.g., a correlation between stellar mass and rotation period).
We can provide mass-dependent disk lifetimes for different disk parameters (e.g., \fig{fig:mdot_init}) and a relation for the SAM after the disk phase that includes the influence of the stellar magnetic field strength and an accretion-powered stellar wind (\equo{eq:jfinal}).
Combining our results with the method presented in \cite{Vasconcelos2017}, could lead to new constraints for stellar and disk parameters (e.g., the mass-dependent distribution of stellar magnetic strength or mass-dependent APSW efficiency).

\section{Conclusion and caveats}\label{sec:final_conclusion}

To investigate the factors that govern the angular momentum distribution of young M~dwarfs, we ran more than 500 long-term (up to 10~Myr) simulations of a rotating star and disk and the interaction between them, i.e., in the innermost disk ($\lesssim$0.05~AU). This large number of simulations, spanning a wide parameter space, was possible due to the implicit solver in the code, which circumvents the limitations of using explicit time steps when simulating disk physics on both small (inner edge) and large (outer edge ) scales. Our model includes physics in a more self-consistent manner (e.g., pressure gradients and magnetic torques in the inner disk, deviations from the Keplerian velocity or the back-reaction in the disk) compared to previous long-term simulations \citep[e.g.,][]{Matt10, Gallet19}.

The main conclusions of this work are:

\begin{itemize}

    \item If the initial accretion rate (a proxy for disk mass) is above a critical value $\Dot{M}_\mathrm{crit}\sim10^{-8}$~$\mathrm{M_\odot\ yr}^{-1}$, the accretion disk effectively "erases" the initial spin of the star and the spin and angular momentum (AM) evolves in a convergent manner that is independent of the initial AM.  In such cases, the final SAM (SAM) is controlled by stellar and disk parameters, i.e., $M_\star$, $B_\star$, $\alpha$, and $W$. Consistent with  \cite{Armitage96}, we find that sufficiently massive disks can regulate stellar rotation.

    \item
    For stellar masses $> 0.3~\mathrm{M_\odot}$, the initial accretion rates $\Dot{M}_\mathrm{init}$ at $\sim 1$~Myr are comparable to $\Dot{M}_\mathrm{crit}$. The SAM of these young M~dwarfs after the disk phase $j_\mathrm{final}$ is independent of the initial conditions and scales with the stellar magnetic field strength as $B_\star^{-1.1}$.
    For stellar masses $\lesssim 0.3~\mathrm{M_\odot}$, $\Dot{M}_\mathrm{init}< \Dot{M}_\mathrm{crit}$ and the initial stellar rotation periods affect the SAM after the disk phase.

    \item
    Stellar dipole field strengths in the range of $\sim 30$~G to $\sim 1$~kG (see \fig{fig:Bcomp}) are necessary to reproduce the nearly mass-independent SAM distribution reported by \cite{Somers17} with higher stellar masses requiring stronger field strengths. These results match the range of observed large-scale dipole field strengths reported in the literature. 
    In addition, the positive relation between the stellar mass and the magnetic field strength is needed to explain the mass-independent SAM distribution in the presence of mass-dependent accretion rates and is suggested by theoretical scaling arguments \citep[][]{Browning16}.

    \item Our scaling relation is also consistent with (but not robustly tested by) the few available observations of SAM, mass, and magnetic field strength in individual stars.
    
\end{itemize}
Our study points to future directions in advancement in both modeling and observations, but should be considered preliminary because of important caveats and limitations of our model:
\textit{Model dimensionality:} Our 1-D model does not resolve the vertical structure of the disk or include vertical gradients as shear. Thus, certain mechanisms such as for example vertical shear instabilities \citep[e.g.,][]{Urpin98, Nelson13, Latter22} and vertically resolved ``dead" zones \citep[e.g.,][]{Jankovic21, Delage21} which could be important for the transport of mass and angular momentum cannot be included. An expansion to 2-D, including the vertical dimension, would allow such phenomena to be included.
\textit{Early disk evolution:} Phenomena relevant to the earlier ($\lesssim 1$~Myr) phase (Class 0 and Class 1 phases) of stellar evolution  of the star-disk system are not  included in the TAPIR code. The effects of a more massive disk, accretion outbursts, and in-falling material from the parent cloud core could be included in future studies. \citet{Vorobyov17c} show that unsteady accretion during this early phase can affect later stellar evolution.
\textit{Pre-main sequence contraction:} In the current TAPIR version the contraction of the central star is described by the simplified Kelvin-Helmholtz contraction.  Due to unsteady accretion, which can occur even in later disk stages \citep[e.g., HBC~722,][]{kospal16}, the contraction can deviate significantly from the Kelvin-Helmholtz relation.  A future improvement could be combining the TAPIR code  with the predictions of a full stellar evolution code \citep[e.g., MESA][]{Praxton19}.
The combination of a stellar evolution code with our current model would further allow a more self-consistent treatment of the evolution of the stellar magnetic field strength, assuming the magnetic energy density scales with the kinetic energy of the convective flow \citep[e.g.,][]{Browning16}.
\textit{Stellar multiplicity:} In the current study, we only consider single stars. Binary or multiple systems tend to influence the evolution of the accretion disk \citep[e.g.,][]{Akeson19, Zagaria21}. The influence of a binary companion on the angular momentum of a star will be studied in future work.
\textit{Transport of mass and angular momentum:} Our model relies on a simple (constant $\alpha$-parameter) prescription for viscous transport of mass and angular momentum in the disk.  $\alpha$ is almost certainly not constant with time and location in the disk, and other mechanisms such as disk winds and photo-evaporation, which cannot be described with this parameterization, are important for disk evolution \citep[e.g.,][]{Konigl11,Roquette21}.  These are currently not included but could, in principle, be added in some future version of our model.

%% file: sections/acknowledgments.tex
\begin{acknowledgements}

The authors thank the anonymous referees for providing comments and suggestions that helped to improve the quality of this work.
The authors thank Ansgar Reiners for the discussion of measurements of stellar magnetic fields.  EG acknowledges a 2021 Ida Pfeiffer Professorship in the Faculty of Earth Sciences, Geography, and Astronomy at the University of Vienna, and support from NSF Astronomy \& Astrophysics grant 1817215.  
 
\end{acknowledgements}

%% file: footer.tex
\bibliographystyle{resources/bibtex/aa}
\bibliography{literature/Tapire,literature/Gaidos}

%% file: sections/appendix.tex
\FloatBarrier

\section{Influence of MEs for different configurations}

\begin{figure}[ht!]
    \centering
         \resizebox{\hsize}{!}{\includegraphics{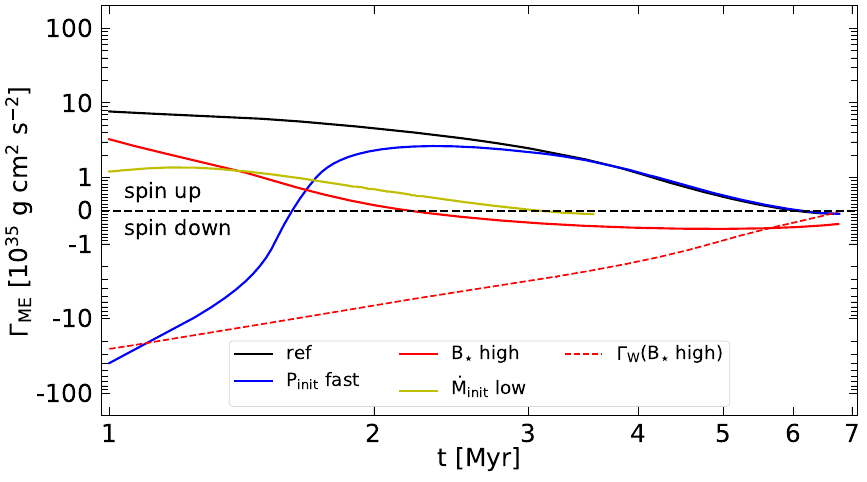}}
    \caption{
    Evolution of the torque due to magnetospheric ejections (MEs) for different configurations (see text).
    The transfer between the spin-up and spin-down regime is marked with a horizontal black dashed line.
    For comparison, the spin-down torque due to the APSW, $\Gamma_\mathrm{W}$, for model \textit{$B_\star$~high} is shown as a dashed red line.
    }
    \label{fig:ME}
\end{figure}

The torque due to magnetospheric ejections (MEs) is shown for different configurations in \fig{fig:ME} compared to a reference model.
The reference model (solid black line) has a magnetic field strength of $B_\star = 0.5$~kG, an initial rotation period of $P_\mathrm{init} = 10$~days, an initial accretion rate of $\Dot{M}_\mathrm{init} = 1\times 10^{-8}~\Msolpyr$, a viscosity parameter $\alpha = 0.01$, and a stellar mass of $M_\star = 0.3~\mathrm{M_\odot}$.
We increase the strength of the magnetic field to $B_\star = 1.0$~kG (model "\textit{$B_\star$~high}", solid red line), decrease the initial rotation period to 1~day (model "\textit{$P_\mathrm{init}$~fast}", solid blue line), and reduce the initial accretion rate to $\Dot{M}_\mathrm{init} = 3\times10^{-9}~\Msolpyr$ (model "\textit{$\Dot{M}_\mathrm{init}$~low}", solid yellow line).
The transfer between the spin-up and spin-down regime is marked with a horizontal black dashed line.
For comparison, the spin-down torque due to the APSW, $\Gamma_\mathrm{W}$, for model \textit{$B_\star$~high} is shown as a dashed red line.

\FloatBarrier

\section{Flat spectrum sources}

Observed accretion rates of flat spectrum sources used in \fig{fig:mdot_init}.

\begin{table*}[ht]
\begin{threeparttable}[t]
\centering
\caption{Stellar masses and accretion rates of flat-spectrum sources with $M_\star<1~\mathrm{M_\odot}$ used in \fig{fig:mdot_init}. In addition, the method for determining the accretion rate (Method$_\mathrm{\Dot{M}_\star}$) is given. If only one reference is given it contains stellar mass and the accretion rate.}              
\begin{tabular}{l c c l l }         
\hline\hline                       
Name & $M_\star~\mathrm{[M_\odot]}$ & $\Dot{M}_\star~[10^{-8}\Msolpyr]$ & Method$_\mathrm{\Dot{M}_\star}$ & Ref.\\
\hline  

V846Ori			&	0.55&	12.0& Ca$_{\rm II}$/Pa$\beta$/Br$\gamma$ lines	 &\cite{Caratti2012}	\\
J05390536-0711052 &	0.55&	13.0&	Ca$_{\rm II}$/Pa$\beta$/Br$\gamma$ lines	&\cite{Caratti2012}	\\
IRAS 05379-0815&			0.73&	7.5&	Ca$_{\rm II}$/Pa$\beta$/Br$\gamma$ lines	&\cite{Caratti2012}	\\
J05413033-0840177&		0.56&	10.0&	Ca$_{\rm II}$/Pa$\beta$/Br$\gamma$ lines	&\cite{Caratti2012}	\\
J03290895+3122562&		0.15& 	1.1& Pa$\beta$/Br$\gamma$ lines	&\cite{Fiorellino21}\\
J03292044+3118342&		0.28&	2.1& Pa$\beta$/Br$\gamma$ lines	&\cite{Fiorellino21}\\
IRAS 04385+2550	&		0.59&	12.0\tnote{a}&		SED fit\tnote{a}&\cite{Guedel2007}, \cite{Ribas2020}\\
DG Tau  	&		0.50&	25.0&	U-band excess &\cite{Guedel2018}, \cite{Alonso2017}\\

\hline       
\end{tabular}
\begin{tablenotes}
\item[a] SED fitting with an artificial neural network using the D’Alessio Irradiated Accretion Disk models \citep[DIAD, see][]{Ribas2020}.
\end{tablenotes}
\end{threeparttable}%
\label{tab:flats}  
\end{table*}

\FloatBarrier

\section{Observational data for stars after the disk phase}
\label{sec:obs_stars}

Stellar observational data used in \fig{fig:Bcomp} and \fig{fig:obs_after} is shown in \tab{tab:obs_stars}.

\begin{table*}[]

\centering
\caption{Stellar observational data used in \fig{fig:Bcomp} and \fig{fig:obs_after}. The respective uncertainties are given when available. Missing data is represented by "---".}              
\begin{tabular}{l c c c c c c c}         
\hline\hline                       
Name & Age~[Myr] & $M_\star$~[$\mathrm{M_\odot}$] & $R_\star~\mathrm{[R_\odot]}$ & $P_\star$~[d] & $B_\mathrm{I}$~[G] & $B_\mathrm{V}$~[G] & $B_\mathrm{QU}$~[G] \\
\hline                                  
    AU~Mic & $23\pm1$\tablefootmark{1} & $0.50\pm0.03$\tablefootmark{2} & $0.75\pm0.03$\tablefootmark{2} & $4.86\pm0.01$\tablefootmark{2} &  $2310\pm200$\tablefootmark{1} & 88\tablefootmark{1} & $2000\pm200$\tablefootmark{1} \\
    WW~Psa & $27\pm7$\tablefootmark{3} & $0.19\pm0.04$\tablefootmark{3} & $0.70\pm0.13$\tablefootmark{3} & $2.37\pm0.01$\tablefootmark{3} & $2750\pm250$\tablefootmark{3} & --- & --- \\
    TX~Psa & $27\pm7$\tablefootmark{3} & $0.11\pm0.03$\tablefootmark{3} & $0.52\pm0.10$\tablefootmark{3} & $1.09\pm0.01$\tablefootmark{3} & $2750\pm250$\tablefootmark{3} & --- & --- \\
    V1124~Ori\tnote{a} & 1.2\tablefootmark{4} & 0.26\tablefootmark{4} & 2.15\tablefootmark{4} & 8.18\tablefootmark{4} & $2090\pm340$\tablefootmark{4} & --- & --- \\
    V1123~Ori\tnote{a} & 0.8\tablefootmark{4} & 0.35\tablefootmark{4} & 2.59\tablefootmark{4} & 7.64\tablefootmark{4} & $2510\pm410$\tablefootmark{4} & --- & --- \\
    V1568~Ori\tnote{a} & 1.3\tablefootmark{4} & 0.40\tablefootmark{4} & 2.37\tablefootmark{4} & 5.54\tablefootmark{4} & $1420\pm230$\tablefootmark{4} & --- & --- \\
    Upp~Sco~55\tnote{b} & 5\tablefootmark{5} & $0.10$\tablefootmark{5} & $0.52$\tablefootmark{5} & $2.90\pm0.30$\tablefootmark{5} & $2300\pm1300$\tablefootmark{5} & --- & --- \\
    TWA 9B & 10\tablefootmark{6} & $0.30$\tablefootmark{6} & $0.91$\tablefootmark{6} & $3.98$\tablefootmark{6} & $3100\pm200$\tablefootmark{6} & --- & --- \\ 
    TWA 8A & $11\pm5$\tablefootmark{7} & $0.45\pm0.10$\tablefootmark{7} & $0.80\pm0.20$\tablefootmark{7} & $4.58\pm0.01$\tablefootmark{7} & $6000\pm500$\tablefootmark{7} & 720\tablefootmark{7} & --- \\ 
    GJ~490B & 35-350\tablefootmark{8} & $0.33$\tablefootmark{9} & $0.34$\tablefootmark{9} & $0.54$\tablefootmark{9} & $3200\pm400$\tablefootmark{9} & 680\tablefootmark{9} & --- \\ 
    V2247~Oph\tnote{a} & 1.0\tablefootmark{10} & 0.35\tablefootmark{10} & 2.00\tablefootmark{10} & 3.40\tablefootmark{10} & $2700\pm100$\tablefootmark{11} & 90\tablefootmark{10} & --- \\
    TWA 7 & $17\pm10$\tablefootmark{12} & $0.62\pm0.03$\tablefootmark{12} & $0.80\pm0.10$\tablefootmark{12} & $5.00\pm0.01$\tablefootmark{12} & $2700\pm200$\tablefootmark{11} & 50\tablefootmark{12} & --- \\
\hline       
\end{tabular}
\begin{tablenotes}
\item[a] Despite the young age of those stars, they are identified as NTTS or weak accretors and thus, included in the list.
\item[b] $P_\star$ calculated from $vsini$ data.
\end{tablenotes}
\tablebib{
(1)~\citet{Kochukhov20}; (2) \citet{Plavchan20}; (3) \citet{Messina17}; (4) \citet{Yang11}; (5) \citet{Reiners09}; (6) \citet{Yang08}; (7) \citet{Hill19}; (8) \citet{Klutsch14}; (9) \citet{Phan-Bao09}; (10) \citet{Donati2010}; (11) \citet{Lavail2019}, (12) \citet{Nicholson2021}.
}

\label{tab:obs_stars}  
\end{table*}

\FloatBarrier